\documentclass[sigconf]{acmart}

\AtBeginDocument{%
  \providecommand\BibTeX{{%
    \normalfont B\kern-0.5em{\scshape i\kern-0.25em b}\kern-0.8em\TeX}}}

\usepackage{xspace}
\usepackage{verbatim}
\usepackage{multirow}
\usepackage{makecell}
\usepackage{soul}

\newcommand{\rv}[1]{#1}

\newcommand{\system}[0]{Graphologue\xspace}

\newcommand{\llm}[0]{LLM\xspace}
\newcommand{\llms}[0]{LLMs\xspace}
\newcommand{\gf}[0]{GPT-4\xspace}
\newcommand{\code}[1]{\texttt{#1}}

\copyrightyear{2023}
\acmYear{2023}
\setcopyright{rightsretained}
\acmConference[UIST '23]{The 36th Annual ACM Symposium on User Interface Software and Technology}{October 29-November 1, 2023}{San Francisco, CA, USA}
\acmBooktitle{The 36th Annual ACM Symposium on User Interface Software and Technology (UIST '23), October 29-November 1, 2023, San Francisco, CA, USA}
\acmDOI{10.1145/3586183.3606737}
\acmISBN{979-8-4007-0132-0/23/10}

\begin{document}

\title[Graphologue: Exploring Large Language Model Responses with Interactive Diagrams]{Graphologue: Exploring Large Language Model Responses\\ with Interactive Diagrams}

\author{Peiling Jiang}
\authornote{Both authors contributed equally to this research.}
\email{peiling@ucsd.edu}
\orcid{0000-0003-4447-0111}
\affiliation{%
  \institution{University of California San Diego}
  \streetaddress{9500 Gilman Dr}
  \city{La Jolla}
  \state{California}
  \country{USA}
  \postcode{92093}
}

\author{Jude Rayan}
\email{jrayan@ucsd.edu}
\orcid{0009-0001-2965-752X}
\authornotemark[1]
\affiliation{%
  \institution{University of California San Diego}
  \streetaddress{9500 Gilman Dr}
  \city{La Jolla}
  \state{California}
  \country{USA}
  \postcode{92093}
}

\author{Steven P. Dow}
\email{spdow@ucsd.edu}
\orcid{0000-0002-1354-9866}
\affiliation{%
  \institution{University of California San Diego}
  \streetaddress{9500 Gilman Dr}
  \city{La Jolla}
  \state{California}
  \country{USA}
  \postcode{92093}
}

\author{Haijun Xia}
\email{haijunxia@ucsd.edu}
\orcid{0000-0002-9425-0881}
\affiliation{%
  \institution{University of California San Diego}
  \streetaddress{9500 Gilman Dr}
  \city{La Jolla}
  \state{California}
  \country{USA}
  \postcode{92093}
}

\renewcommand{\shortauthors}{Jiang et al.}

%%
% Abstract.
\begin{abstract}
Large language models (LLMs) have recently soared in popularity due to their ease of access and the \rv{unprecedented ability to synthesize text responses to diverse user questions}. However, LLMs like ChatGPT present significant limitations in supporting complex information tasks due to the insufficient affordances of the text-based medium and linear conversational structure. Through a formative study with ten participants, we found that LLM interfaces often present long-winded responses, making it difficult for people to quickly comprehend and interact flexibly with various pieces of information, particularly during more complex tasks. We present Graphologue, an interactive system that converts text-based responses from LLMs into graphical diagrams to facilitate information-seeking and question-answering tasks. Graphologue employs novel prompting strategies and interface designs to extract entities and relationships from LLM responses and constructs node-link diagrams in real-time. Further, users can interact with the diagrams to flexibly adjust the graphical presentation and to submit context-specific prompts to obtain more information. Utilizing diagrams, Graphologue enables graphical, non-linear dialogues between humans and LLMs, facilitating information exploration, organization, and comprehension.

\end{abstract}

%%
% ACM Computing Classification System.


%%
% Keywords.

%% Teaser.
\begin{teaserfigure}
  \includegraphics[width=\textwidth]{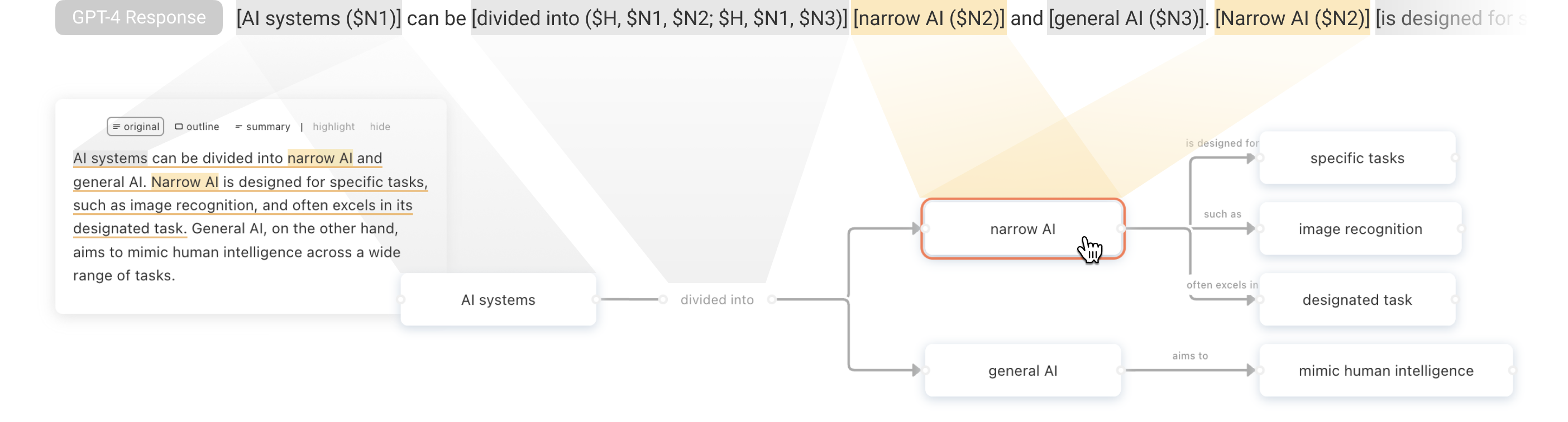}
  \caption{\rv{\mbox{\system} constructs an interactive diagram in real-time as GPT-4 text responses are streamed in.}}
  \Description{\mbox{\system} constructs an interactive diagram in real-time as GPT-4 text responses are streamed in.}
  \label{fig:teaser}
\end{teaserfigure}

\maketitle

\section{Introduction}
\label{sec:introduction}

\rv{\mbox{\textbf{L}}arge \mbox{\textbf{L}}anguage \mbox{\textbf{M}}odels (LLMs) have seen a surge in popularity due to their impressive ability to generate high-quality textual responses to natural language prompts across a wide variety of tasks~\mbox{\cite{sparks2023}}. More than a billion people have used interfaces to LLMs, such as ChatGPT, to obtain information and answers to questions. With the potential to dramatically transform how people complete information processing tasks, LLMs are becoming increasingly important tools in various fields~\mbox{\cite{gpt4technical, sparks2023}}.}

Despite their potential, interactions with LLMs are primarily mediated through text-based conversational interfaces, which presents inherent limitations in terms of supporting complex information activities. As a sequence of symbols, text can be insufficient for communicating concepts that contain complex relationships and structures, often leading to verbose responses that demand substantial effort to digest \cite{chang2002effect}. In addition, the linear conversation structure can hinder the iterative concept exploration workflows that employ non-linear structures (e.g., brainstorming ideas), resulting in excessive and verbose conversational exchanges where the users often lose track~\cite{goodwin2015professional}. These intrinsic constraints of text-based conversational interfaces limit the effectiveness of leveraging \llms for complex information tasks~\cite{qin2023chatgpt}.

Graphical representations, such as diagrams and charts, on the other hand, can compensate for the aforementioned limitations by displaying information in a non-linear manner, enabling the more flexible organization of concepts and reducing the cognitive load needed for comprehension \cite{ainsworth2003effects, wheeldon2011picture, larkin1987diagram}. Interactive graphics can further facilitate the manipulation of information, enabling users to effectively obtain, organize, transform, and make sense of information \cite{conceptmapmobile2011,fatemeh2011icmap,hwang2013concept,wu2012innovative}. For these reasons, graphical representations have been extensively studied and utilized in various fields, including HCI \cite{guastello1989verbal}, Visualization \cite{chen2010information}, Cognitive Science \cite{cheng2001cognitive}, Communication \cite{moura2017learning}, and beyond. The goal of this project, therefore, is to capitalize on the advantages of graphical representations to mitigate challenges associated with text-based conversational interfaces in LLM applications. We envision a graphical conversation between humans and LLMs, continuing the fruitful and long-lasting endeavor 
in HCI~\cite{sketchpadsutherland1964}.

Because text and graphics are both versatile media that can utilize different formats and styles for different tasks, it is important to target specific tasks for meaningful generalization. We focus on supporting exploratory information-seeking, concept explanation, and question-answering tasks with LLMs using graphical representations. To understand the challenges of utilizing LLMs for these tasks, we conducted a formative study with ten participants to observe how they used ChatGPT to explore and learn about a domain of interest. Participants reported that the text-based responses from LLMs are often verbose and time-consuming to comprehend. In addition, we observed that the text-based linear conversational structure imposed many cumbersome interactions, such as repetitive copy-and-paste and back-and-forth scrolling, to carry out complex information tasks.

Informed by the formative study, we designed \system{}, an interactive system that converts textual responses from LLMs into graphical diagrams, in real-time, to facilitate complex and multi-faceted information-seeking and question-answering tasks. \system{} employs novel prompting strategies to have LLMs recognize and annotate entities and relationships inline within the generated responses to facilitate the real-time construction of node-link diagrams. To avoid overly complex diagrams, \system{} employs prompting strategies and interaction techniques to enable users to flexibly control the complexity of the diagrams. For example, users can toggle the diagrams to show only salient relationships, collapse branches of the diagrams to reduce the presented information, and combine separate smaller diagrams into one diagram to view all concepts as a whole. To gain more information about concepts presented in the diagram, users can employ direct manipulation of the graphical interface, which is subsequently translated into context-aware prompts for the LLM, enabling users to engage in a ``graphical dialogue'' with \llms.

To evaluate how effectively \system facilitates graphical interaction with LLMs for information-seeking tasks, we conducted a user evaluation with seven experienced LLM users. We found that \system{} helped them quickly grasp key concepts and their connections while ensuring sufficient control of the complexity of the diagrams. Together with other representations that are interactively synchronized with the diagrams, such as the raw text and the outline, \system{} enabled participants to leverage the combined strengths of different representations to understand LLM responses at various levels and scales.
This work thus makes the following contributions:

\begin{enumerate}
    \item A formative study that uncovered limitations in using a conversational interface for complex sensemaking tasks.
    \item \system{}\footnote{The prototype is available at \url{https://graphologue.app}.}, an interactive system that employs prompting strategies and interaction techniques to enable real-time construction of and interaction with node-link diagrams based on LLM-generated information to facilitate comprehension and exploration.
    \item A qualitative evaluation that provided insights regarding the advantages and disadvantages of the diagrammatic representation, the complimentary usage of multiple representations, and future directions of employing graphical interfaces to interact with LLMs.
\end{enumerate}

\section{Related Work}
\label{sec:related}
As our research aims to address the many challenges of natural language interfaces for LLMs by using graphical representations, we review prior work on Natural Language User Interfaces and LLMs, generating graphical representations from text, as well as Visualization and multilevel abstraction of information.

\subsection{Natural Language User Interfaces and LLMs}
The impressive performance and the public release of LLMs have sparked imaginations for applying this technology to various domains such as programming~\cite{chen2021evaluating, pearce2022examining, evallmcode2022}, writing support~\cite{yuan2022wordcraft,gero2022sparks,chung2022talebrush}, learning~\cite{macneil2022automatically,kung2023performance,baidoo2023education}, and many others~\cite{singh2022progprompt,borsos2022audiolm,chang2023muse,di22idea}, and further promoting the notion of natural language interfaces the HCI community has been exploring.

Natural language interfaces offer the key benefit of enabling users to directly articulate their intended actions and goals without learning and utilizing complex manual user interfaces. Pioneering systems, such as SHRDLU~\cite{winograd1972shrdlu}, Put-that-there~\cite{bolt1980put}, and Quickset~\cite{cohen1997quickset} allowed users to verbally instruct a computer with natural language commands. Later research extended this interaction paradigm to data analysis \cite{datatone}, image editing \cite{voicecuts}, and more. A limitation of these systems, however, is that they require users to translate their high-level design intention to low-level, rigid commands or queries, limiting the fluidity and expressiveness afforded by language as a communication medium~\cite{crosspower2020}. 

Another approach to natural language interfaces has been extracting user intents from their natural expressions to enable less rigid communication between humans and computers by leveraging advanced natural language understanding and domain-specific knowledge. Iris, for example, enables users to describe data analysis goals and disambiguate system interpretations using natural expressions \cite{iris2018}. CrossData infers the desired data values and operations from text to report in the data insights without instructing the system~\cite{crossdata2022}. Crosspower employs a human-in-the-loop approach by enabling users to interact with linguistic structures in a video script to convey high-level design goals regarding the graphical content and structures \cite{crosspower2020}. 

While the unparalleled language understanding and generation capability of LLMs enables users to obtain meaningful responses with flexible natural language expressions, the inherent intelligence architecture of LLMs poses additional usability challenges. Many tasks require users to go through arduous and time-consuming prompt engineering to produce well-crafted prompts, thereby ensuring results that align with their intents~\cite{promptingvizlang2022, reynolds2021prompt, designguideprompt2022}.

Our work leverages advances in Natural Language Processing (NLP) but shares the spirit of Sketchpad, a seminal work in HCI that pioneered the graphical communication between humans and machines~\cite{sketchpadsutherland1964}. Specifically, we leverage the generative power of GPT-4 to not only obtain information but also to annotate its own text-based LLM responses to facilitate the simultaneous creation of graphical representations.

\subsection{Text to Graphical Representations}

Graphical representations are prevalent across various domains to enhance communication and sensemaking \cite{baker2009using}. By leveraging human aptitude for visual information processing, they offer advantages in comprehension, memory, and inference of the content \cite{agrawala2011design}, making them an effective tool to improve information understanding and learning \cite{linkgoals&resources2005,CNT2004,conceptmapmobile2011,booc.io,jiang2023log}. In addition, the generation and modification of graphical representations, such as sketching and annotating, make them ideal for sensemaking tasks \cite{Conceptmaps2005,goodwin2015professional,palani2022interweave,liu2022wig}. Consequently, significant research in HCI and visualization has explored the design and creation of graphical representations. The recent advancement of NLP has further enabled the automatic generation of graphical content such as visualizations~\cite{NL4DV2020}, 3D scenes~\cite{chang2015text}, animations~\cite{hong2022avatarclip}, and videos~\cite{esser2023structure}.

For example, systems have been developed to generate visualizations and link existing ones with natural language descriptions to assist the communication and comprehension of data insights  \cite{couplingstory2viz2018, linkingtxt&tables2018, elasticdocs2018, Text-to-viz2019}. Techniques have been proposed to automatically generate videos from natural descriptions \cite{singer2022makeavideo}, structured markdown documents \cite{markdownvideo2021}, or informal conversations \cite{crosscast2020} to create a visual consumption experience without significant manual effort.

Closely related is work that explores the generation of node-link diagrams based on the text from a variety of sources such as video transcripts \cite{conceptguide2021,conceptscape2018,videoconceptmapgen2020}, documents \cite{evoq2018}, and social media data \cite{socialmediatextviz2017}. For example, More et al. leveraged NLP to generate Unified Modeling Language (UML) diagrams from natural language specifications to facilitate the analysis of software systems \cite {more2012generating}. ConceptGuide compiles the transcripts from multiple YouTube videos of a certain topic and then constructs a concept map revealing the various relationships between the videos in order to ease the video-based learning process \cite{conceptguide2021,conceptscape2018}.

Unlike previous work that generates diagrams using existing static text content, this work explores the interaction with node-link diagrams generated from the dynamic text output from LLMs. We explore prompting strategies that enable real-time construction of and interaction with the diagrams to facilitate the comprehension of information provided by LLMs.

\subsection{Multilevel Abstraction and Visualization of Information}
Extensive research in HCI and Visualization has investigated interaction and visualization techniques that allow users to quickly grasp an overview of complex information while maintaining access to detailed, low-level information or system functionality.

In the field of information visualization, the `focus + context' design principle states that users need both detail and overview to make sense of information~\cite{card1999}. Information of interest and importance should be displayed in detail, while relevant context should be presented simultaneously to show how these informational details connect to the context.

Bederson and Hollan introduced semantic zooming, which displays information at varying levels of detail corresponding to the scale within a zoomable user interface~\cite{pad++1994}. Norman et al. proposed the concept of progressive disclosure, suggesting that interfaces should progressively inform users about a system by gradually providing pieces of information that contribute to the overall understanding~\cite{normancentered1986}.
Xia et al. explored how users could leverage flexible representational transformation to adjust content representations semantically, structurally, and temporally according to their needs, rather than conforming to a single representation imposed by the user interface~\cite{writlarge2017}. This concept was later applied to a program visualization system, allowing programmers to visually inspect program behaviors at different levels of scope and abstraction~\cite{crosscode}. Victor explored how varying levels of abstraction over data, procedures, and iteration could help explain complex system behaviors~\cite{victor2011abstraction}.

\system{} builds upon these prior works to ensure the graphical diagrams are easy to understand by enabling users to flexibly control the levels of detail of the diagrams and synchronizing them with the original text, which provides the full context.

\section{Formative Study}
We conducted a formative study aiming to uncover the prevailing experiences and challenges associated with using current conversational interfaces to interact with LLMs. The results of this study informed the design choices we made for \system.

\subsection{Participants and Procedure}

Ten participants with a variety of ChatGPT experiences were recruited, including two first-time users, six casual users who are familiar with ChatGPT, and two experienced users who use it daily with advanced prompting techniques and have developed applications using OpenAI's API. The study sessions were conducted over Zoom for an hour each, and participants received 15 USD as compensation for their participation.

Participants completed a pre-task survey that collected demographic information and their experience with ChatGPT. They were then asked to select one topic (from Neuro-divergence, Supply and Demand, Northern Lights, and Inflation) to explore using ChatGPT. Participants were provided with a task description document containing questions related to the chosen topic, which were designed to help them broadly and deeply explore the topic. Participants were given 30 minutes to explore the topic and then interviewed to reflect on their experiences with a focus on the usability pain points of using ChatGPT to obtain, manage, and understand information. Participants were also encouraged to share functionality that they thought would help circumvent the issues they had encountered. The interviews were recorded, transcribed, and analyzed using the reflexive thematic analysis method~\cite{braun2012thematic}. 

\subsection{Findings and Discussion}

All participants explored the concepts and questions mentioned in their assigned task descriptions, engaging in an average of nine conversational exchanges with ChatGPT. We present the key themes of the \textbf{C}hallenges that emerged from the interviews.\newline

\noindent\textbf{C1. Response Content is Verbose and Lacks Structure.}
Participants expressed concerns over ChatGPT's explanations and commented that \textit{``it was definitely very easy to get overwhelmed by information thrown at [them] sometimes''} (P2). Even if the questions were intentionally framed with a specific scope for short responses, ChatGPT provided long answers (P5). If participants sensed that ChatGPT was generating a redundant response with extraneous background information, they tended to click `Stop generating' button (P3). When participants asked a follow-up question, many made similar comments that \textit{``ChatGPT tends to repeat the whole thing, and [they] would have to skim over some stuff that [they] already know''} (P5). It was clear that \textit{``ChatGPT was trying to be as exhaustive as possible answering [their] questions''} (P2).

Regarding the format of the information presented, P3 found bullet-point content was easier to understand. P5 commented that \textit{``there's not really a visual hierarchy in the text,''} making it difficult to navigate a large amount of text. Participants suggested functionalities to circumvent these issues, including being able to see \textit{``different formats''} (P3), having information management capabilities, like collapse (P6), and shortening parts to prevent repetition (P5).\newline

\noindent\textbf{C2. Lack of Flexible Interaction with the Response Text.}
All participants extracted some parts of the ChatGPT responses and tried to query them as a prompt to explore them further. However, they expressed the desire to interact with the response directly rather than through a series of conversations. For example, P2 wished they could highlight a part of the ChatGPT response and directly ask \textit{``Oh, what does this line mean?''}. If participants were unsure of certain parts of the response, they had to manually write it down, or copy-paste from the original text, and prompt each question one by one to get clarification, which is time-consuming and \textit{``increases [their] mental load''}~(P9).\newline

\noindent\textbf{C3. Lack of Organization Across Multiple Responses.}
Keeping track of the various questions and answers previously encountered was found challenging for all participants, as P1 noted that they \textit{``struggled a little bit to like remember everything.''} This was further exacerbated by \textit{``redundant answers [that] lack organization''} and form a single stream of questions and answers (P3).

Almost all participants wanted the ability to organize the information collected during the multiple back-and-forth exchanges. Due to a lack of organization,  P5 had to \textit{``scroll through a lot''} to find relevant information in a previous response, and suggested a bookmarking technique that would enable them to \textit{``annotate stuff that you want to go back to later.''} P2 suggested providing an overview of what information has been explored with bullet points. Alternatively, P2 and P4 recommended organizing the responses spatially in a mind-map form that visualizes \textit{``the connections or the relations between the questions [they] ask,''} which will be \textit{``very useful in terms of understanding the whole topic''} (P2).

\subsection{Summary}

The formative study reveals that the participants found several challenges with respect to the quantity, organization, and presentation of, and interaction with the ChatGPT responses. They found the responses to be verbose, making them difficult to track, process, and comprehend. The linear conversation contributes to the disorganization of the information embedded in a series of exchanges with ChatGPT. Additionally, there is no direct interactive control over textual responses, which makes it hard for users to specify their intent in follow-up prompts. Overall, the findings indicate that a better representation of information is needed to enable intuitive understanding and flexible exploration of information from LLMs.

\section{Design Goals and Rationale}
\label{sec:design}

Based on findings from the formative study and the iterative prototyping and evaluation process, we derive four design goals for a diagram-oriented interaction with LLMs. We first describe the \textbf{D}esign goals, and then a scenario (Section~\ref{sec:scenario}) to ground how these goals can be supported by a novel system and lead to fluid interaction with LLM-generated information.\newline

\noindent\textbf{D1. Diagram as Entry Point.} Our goal was to use diagrams to facilitate the comprehension of LLM responses. However, we found first presenting text from LLMs, and then displaying the diagrams generated from the text, increases rather than decreases the cognitive effort. \rv{This is because LLMs, especially \mbox{\gf}, take significant time to generate complete and comprehensive responses \mbox{(\textbf{C1})}.}
Users read the responses in real-time as the text comes out, and therefore, presenting full diagrams subsequently requires extra time to process. Moreover, the diagrams may not match users' preconceived mental picture, imposing an additional burden on the short-term memory to align the concepts absorbed through reading with diagrams that come afterward.
\rv{Prior research has also explored how different types of visuals, such as diagrams and charts, can serve as simultaneous and preferred facilitators for various text-oriented tasks, including reading and chatting \mbox{\cite{stokes2022text,stokes2023striking,r40,jian2015using,laban2021simple,clark1991dual}}.}

Therefore, we propose generating diagrams \emph{concurrently} with the text and ensuring diagrams are the entry point to the LLM-generated information to aid comprehension.\newline

\noindent\textbf{D2. Flexible Control of Diagram Complexity.} Diagrams can be difficult to understand if they are overly complicated. Therefore, it is essential to manage the complexity of diagrams. The perceived complexity of a diagram can come from two sources: \rv{the amount of information in the original text \mbox{(\textbf{C1})}} and the presentation mechanism of the diagrams. To avoid overwhelming the users with complexity, the users should be able to flexibly control the amount of information to be visualized in the diagram and how the available information should be revealed.\newline

\noindent\textbf{D3. Diagram-Based Exploration.} By utilizing diagrams as the main interface with LLMs, \rv{typical information tasks should be supported through interaction with the nodes and links in these diagrams \mbox{(\textbf{C2})}, which has been proven effective in improving information tasks \mbox{\cite{spoerri1993infocrystal,kim1999why}}}. Users should be able to interact with the diagram to acquire more information, such as further exploring an unfamiliar concept by requesting more explanations or examples. Similarly, users should be able to collapse or trim parts of the diagrams if they are irrelevant to their goal. \rv{Explorations beyond the initial prompt and response should be organized through expanding and trimming of the diagrams \mbox{(\textbf{C3})}.}\newline

\noindent\textbf{D4. Synchronized Interaction Between Diagrams and Text.} From our informal user tests during system development, we found users often refer back to the original text for two reasons. Firstly, although node-link diagrams help users quickly grasp the main concepts and connections, they may need to consult the original text for details about specific concepts or relationships they find intriguing or challenging to understand from the diagrams. Secondly, since we use \gf to identify entities and relationships for diagram construction, occasional recognition errors can result in inaccurate visualizations. In these cases, users rely on the original text to verify their understanding and to correct misconceptions.
\rv{Therefore, we propose that the text and diagrams remain synchronized, allowing users to easily locate relevant text from the diagrams and vice versa to leverage the combined strengths of different representations and ensure an efficient information-processing experience\mbox{~\cite{AINSWORTH2006183}}.}

\section{Envisioned Scenario}
\label{sec:scenario}

We describe a scenario that illustrates the workflow of \system{}, a system to support the above design goals for exploratory information seeking.

While working on a deadline, Margaret felt a tremor and confirmed a 4.5 magnitude earthquake online. Living in an earthquake zone but lacking knowledge about them, she wanted to use ChatGPT, a tool that she had been using lately, to learn more about earthquakes. However, weary from hours of writing, she preferred a quicker way to understand the topic. She recently heard about \system, a tool that converts LLM text responses into diagrams for easier comprehension, and decided to give it a shot.

In \system, she started by typing `What is an earthquake?'. As response text streamed into the interface from the LLM, she noticed a node-link diagram was being constructed piece by piece on the side simultaneously (\textbf{D1}). By glancing at the diagram, she quickly understood that `tectonic plates' - (`shift along') - `faulty line' and - (`generates') - `seismic waves.' As a scientist, she wanted to ensure the logical relationship was correct. When she pointed to `seismic waves,' she noticed the corresponding text was highlighted, allowing her to easily refer to the original text for full details (\textbf{D4}).

In the next diagram, she saw `Richter Scale,' which measures the intensity of earthquakes, and wanted to know the earthquakes that measured `Magnitude 7.' She clicked on the `Magnitude 7' node and then the `Examples' button, and noticed more connections and nodes forming from the node, listing three prior earthquakes, including the `2010 Haiti earthquake' (\textbf{D3}). Seeing that, she recalled the devastating damages of that earthquake that she saw on TV.

Each paragraph and its corresponding diagram explain one particular aspect of the earthquake, allowing her to gain information about the earthquake piece by piece (\textbf{D2}). After she went through all these diagrams, she wanted to know how these concepts are related together as a whole. She switched to the ``merged diagram'' view, and immediately she saw all the smaller diagrams animate and combine into one diagram showing the complete picture (\textbf{D2}). She then said to herself, ``This is groundbreaking.''
 
\begin{figure*}[ht]
    \centering
    \includegraphics[width=\textwidth]{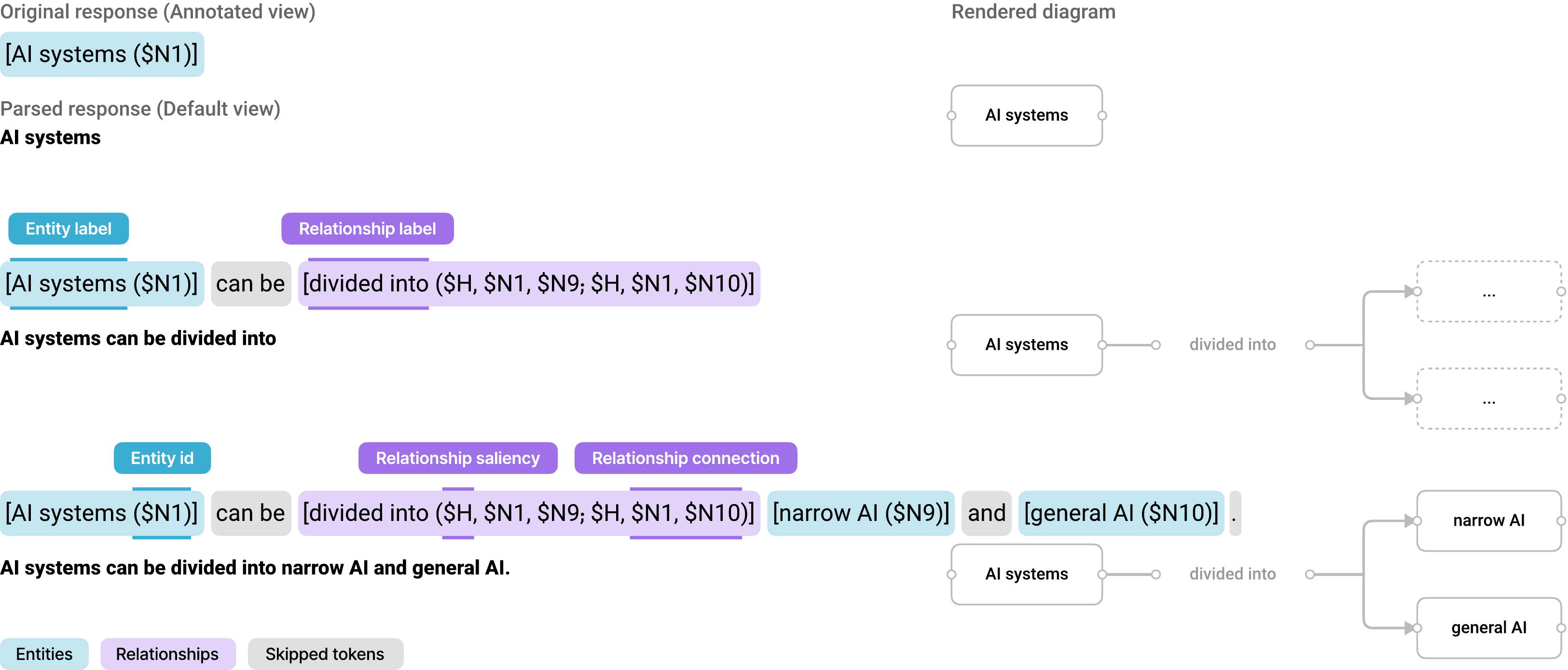}
    \caption{As \gf responses stream in, \system parses them in real-time, removes inline annotations for the interface, extracts entities and relationships, and constructs the corresponding diagrams.}
    \Description{As \gf responses stream in, \system parses them in real-time, removes inline annotations for the interface, extracts entities and relationships, and constructs the corresponding diagrams.}
    \label{fig:ext}
\end{figure*}

\section{Prompting for Diagram Generation}

A primary design goal for \system is to have diagrams as the entry point for people to receive information from \llms (D1). The common prompt chaining strategy, however, requires additional rounds of processing and leads to increased waiting time for the user~\cite{wu2022promptchainer,wu2022ai}. To enable the diagrams to be constructed simultaneously as the response streams in, we iteratively develop our prompts to have the \llm annotate the entities and relationships inline with the tokens. This enables \system{} to provide both the text responses and the diagrams at the same time.

We develop and test the following prompting strategies with OpenAI's \gf, the most advanced and publicly available \llm to date. A full list of original prompts can be found in Appendix~\ref{app:p}.

\subsection{Diagram Construction}

We outline our key prompt components, which work together to instruct \gf to generate an initial response that facilitates dynamic diagram construction (D1) and enables easy control over the complexity of presented information (D2) through interactive diagrams (Appendix~\ref{app:p-initial}).

\subsubsection{Dividing Responses into Paragraphs}

Smaller and more manageable diagrams, based on portions of the response, effectively prevent users from being overwhelmed by an excessive number of nodes and connections in a single diagram derived from the full response. To achieve this, we instruct \gf to structure its initial response into separate paragraphs, each focusing on a single theme, aspect, or topic, and corresponding with one diagram (Figure~\ref{fig:system}.b).
Interacting with separate diagrams allows users to navigate through different sections of the response more easily, resulting in reduced information overload and improved comprehension.

\subsubsection{Annotating Entities}

\gf is instructed to annotate entities in the text to serve as \emph{nodes} in the diagrams. While entity labeling and co-reference resolution are classic tasks in NLP, traditional NLP techniques struggle to perform well without complete sentences \cite{piskorski2013information}. However, we found \gf excels in annotating the entities \emph{simultaneously during the text generation}, as shown in Figure~\ref{fig:ext}.

We instruct \gf to assign a unique identifier (such as \code{\$N1} in \code{[Artificial Intelligence (AI) (\$N1)]}) for each entity referring to the same concept, i.e., co-reference, to enable entities to be associated by the relationships. \gf has surprisingly superior co-reference capabilities and manages identifiers automatically, avoiding repetition and ensuring consistent labeling throughout the response. As shown in the example paragraph A in Appendix~\ref{app:p-initial}, \code{Artificial Intelligence (AI)}, \code{AI systems}, and \code{It} were all co-referenced and identified as \code{\$N1}.

\subsubsection{Annotating Relationships and Saliency}

\begin{figure*}[ht]
    \centering
    \includegraphics[width=\textwidth]{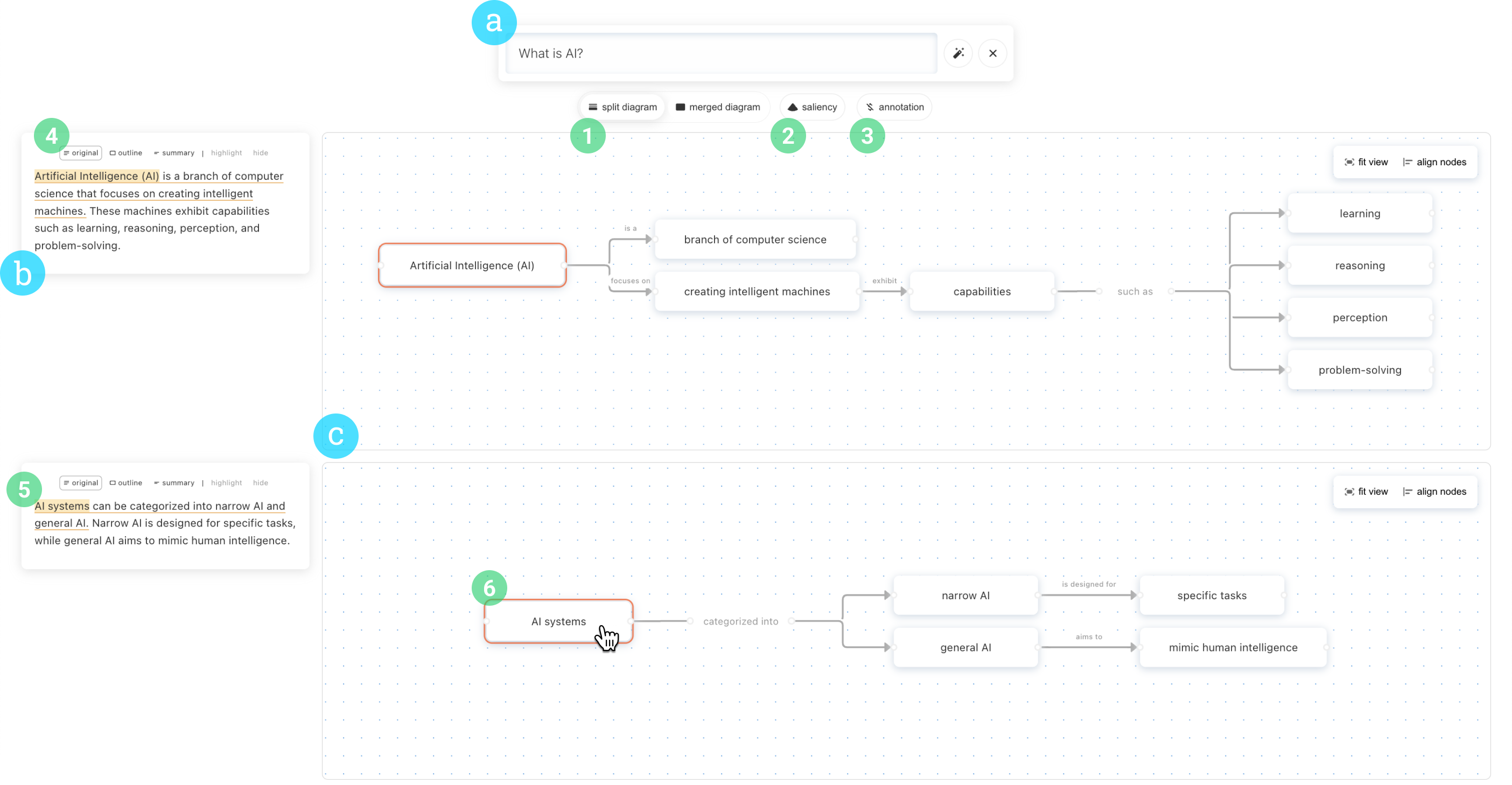}
    \caption{The \system interface, including the question input box (a), text response blocks (b), and the diagrams (c). The user can switch between the default split-diagram and merged-diagram views (1), modify the saliency filter (to show all or only high saliency ones) (2), toggle showing the raw \gf response with annotations or the parsed text (3), and change the text block display between Original, Outline, and Summary (4). When a user hovers on a node in the diagram (6), all co-referenced nodes and their corresponding text tokens are highlighted (5).}
    \Description{The \system interface, including the question input box (a), text response blocks (b), and the diagrams (c). The user can switch between the default split-diagram and merged-diagram views (1), modify the saliency filter (to show all or only high saliency ones) (2), toggle showing the raw \gf response with annotations or the parsed text (3), and change the text block display between Original, Outline, and Summary (4). When a user hovers on a node in the diagram (6), all co-referenced nodes and their corresponding text tokens are highlighted (5).}
    \label{fig:system}
\end{figure*}

In addition to entities, we prompt \gf to identify and annotate relationships between these entities inline (Figure~\ref{fig:ext}). These relationships serve as the \emph{links} in the node-link diagram. A single relationship described in the text may encompass multiple connections between different pairs of entities, and \gf is instructed to include them in a single annotation, which we utilize to organize the connections together when rendering them on the canvas (like ``such as'' in Figure~\ref{fig:system}.c). \gf demonstrates an impressive ability to identify nearly all relationships and associate the corresponding entities with annotations.

However, annotating all relationships can result in the inclusion of less important ones in the text, which may hinder users from grasping the main idea and nature of the original response. Consequently, this may lead to information overload and clutter when rendered as a diagram. To avoid this, we leverage saliency filters to manage the complexity of the rendered diagram. We instruct \gf to annotate each relationship pair with saliency levels for the connections, designated as either \emph{high} (\code{\$H}) or \emph{low} (\code{\$L}). By default, we render the diagram with only the high-saliency relationships to mitigate clutter, while the user can adjust the saliency level control on the interface to show all relationships.

One intriguing aspect of relationship annotations is \gf's ability to \emph{prospectively} annotate the involved entities. Even before an entity has been mentioned and labeled in the response, the connecting relationship annotation already incorporates its identifier (Figure~\ref{fig:ext}). This allows us to add new nodes to the existing diagram structure before the corresponding node entities have finished streaming in. Instead, a placeholder node will appear to help draw the user's attention, enabling a more fine-grained and responsive unveiling of the dynamically constructed diagrams.

\subsection{Error Prevention and Correction}
\label{sec:correction}

\gf demonstrates powerful inline annotation capabilities. Nevertheless, it is not perfect and occasionally makes mistakes. We develop our prompts and interactions of \system to prevent and correct the errors in \gf annotations.

\subsubsection{Avoiding Recurring Problems}
As we develop the prompting rules mentioned above, we identify several common errors made by \gf, which resulted in impaired response time, redundant information, non-parsable responses, or incorrect diagrams. These problems include assistant-style responses (e.g., ``Sure, I can help you with it...''), inconsistent annotation formats (e.g., missing square brackets), misidentifying conjunctive adverbs as entities (e.g., ``therefore'' or ``since then''), and repeating tokens inside and outside of the annotation (e.g., \code{narrow AI [narrow AI (\$N9)] }). We iteratively tune our prompts to explicitly avoid these behaviors and instruct \gf to respond concisely and consistently. 

\subsubsection{Self-Correction with Additional Rounds of Processing}

Two other types of annotation errors in recognizing entities and relationships are addressed through additional rounds of prompting: \emph{dead-end relationships}, which involve annotating relationships that connect non-existent entities; and \emph{orphan entities}, which include annotating entities that are not involved in any relationships within the rest of the response. These mistakes result in empty nodes (nodes with no labels, showing ``...'' instead) and orphan nodes (nodes disconnected from the rest of the diagram), which hinder the user's ability to follow and digest the diagram for information comprehension. To address these issues, we identify such errors upon completing each paragraph and prompt \gf to self-correct these inconsistencies, asking for a corrected version of each sentence that contains errors one by one.

In the self-correction prompt, the previously generated paragraph serves as context. We pinpoint the specific sentence requiring correction and dynamically describe the issue, e.g., ``entities labeled \code{\$N11} and \code{\$N12} are mentioned but lack connecting relationships.'' We instruct \gf to re-annotate the sentence or slightly rewrite it if needed for better annotation (Appendix~\ref{app:p-corr}).

The correction process is done in parallel, and other interactions with the diagram remain unblocked. Once an updated annotation is complete, the diagram is adjusted and animated to reflect the changes. We found that \gf is able to improve the annotation and correct many errors when asked again with issues directly pointed out \cite{shinn2023reflexion}, and we allow users to modify the graph further as needed, which we introduce in Section~\ref{sec:user-corr}.

\subsection{Information Quantity Adjustment}

We limit the initial response from \gf to fewer than four paragraphs, each containing around 2--3 sentences. This would allow the users to use simple diagrams as the starting point of the interactive exploration process without being overwhelmed by complex diagrams. From the initial diagrams, however, the amount of information can be further adjusted, resulting in response text and diagrams of varying lengths and levels of complexity, allowing users to flexibly reduce the amount of information or delve into more detailed and elaborate content as desired (D2).

\subsubsection{Summarizing Each Paragraph of the Response}

\begin{figure*}[ht]
    \centering
    \includegraphics[width=\textwidth]{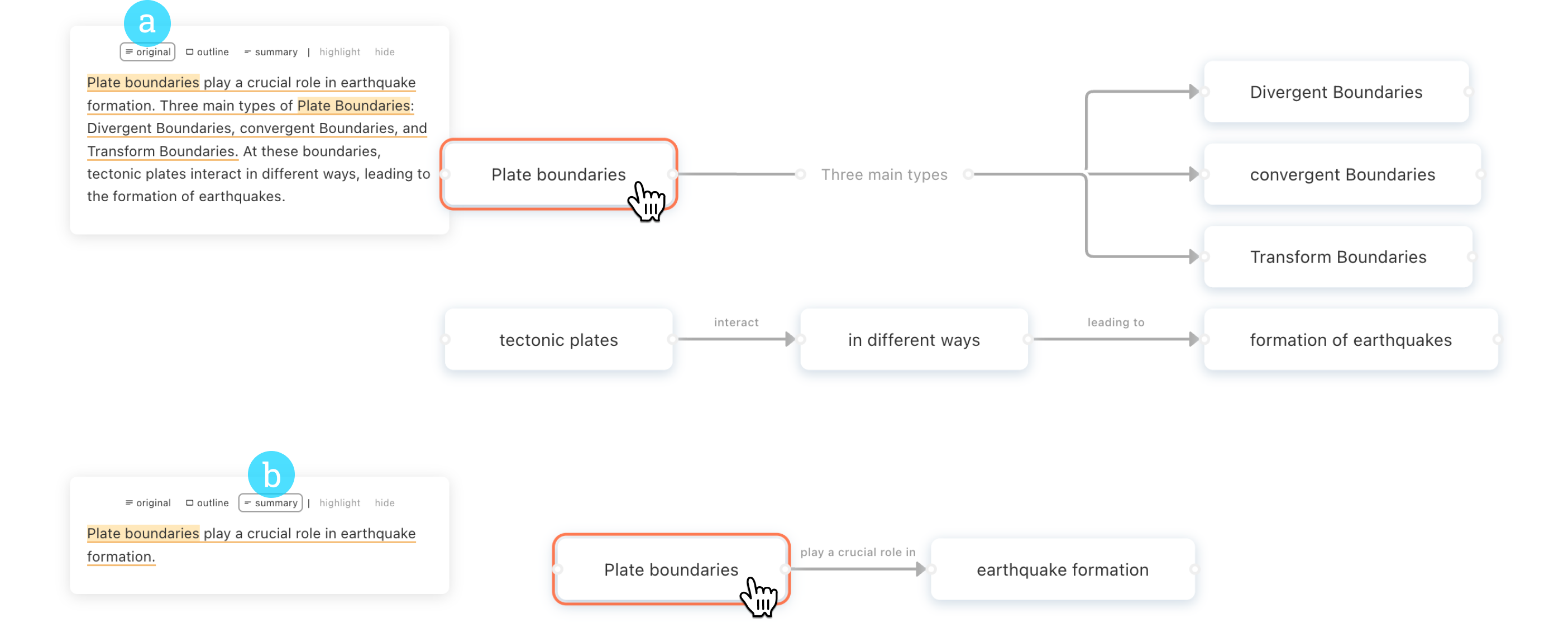}
    \caption{\rv{The original response (a) and its summary (b) from \mbox{\gf}, with the corresponding diagrams.}}
    \Description{The original response (a) and its summary (b) from \gf, with the corresponding diagrams.}
    \label{fig:summary}
\end{figure*}

Saliency filtering helps prevent overwhelming users with excessive information, but sometimes they may want to quickly grasp the most important idea of each paragraph for a brief understanding of the response. To enable this, we prompt \gf to generate a short, one-sentence summary for each paragraph immediately after it is completed. This captures the key concept, and the corresponding diagram is simplified to include only 3--5 nodes, allowing for a concise and easily digestible view. (Figure~\ref{fig:summary}).

Maintaining the identifiers of entities present in the original text and the summary is crucial for enabling smooth diagram transitions between different levels of complexity and retaining context for users to digest. To achieve this, we provide \gf with the annotated response when prompting for summaries. This approach ensures that the same entities are matched across different text and diagrams, allowing for synchronized interactions between the diagram and the text, as introduced in Section~\ref{sec:sync}.

\subsubsection{Asking for More Details}

Being able to customize the content based on individual needs and interests facilitates engagement and in-depth understanding at varied levels and scales~\cite{brusilovsky2001adaptive}. We allow users to request more information about the response, including expanding specific paragraphs of the response (Figure~\ref{fig:hover}.b) with additional explanations or examples, as well as introducing new paragraphs and corresponding diagrams that explore additional aspects of the subject matter (Figure~\ref{fig:hover}.c).

\section{\system{} Interface}
\label{sec:system}

\system transforms textual responses from \llm into interactive diagrams, utilizing entity and relationship annotations streamed and interleaved with the original response. Below, we describe the rich interactions powered by \llm and \system's interface.

\subsection{Constructing the Interactive Diagrams}

To enable users to utilize the node-link diagrams as the entry point for their comprehension, it is essential to have these diagrams responsively updated as new entities and relationships are identified. To accomplish this, we parse the text streamed in from \gf and immediately add new entities and relationships to diagrams as they are generated. This ensures that users have a responsive and up-to-date visual representation of the LLM-generated information.

The diagrams are designed to closely reflect the underlying annotated text. Several design choices are made to facilitate this goal. For example, when multiple entities are extracted with the same identifier through co-reference understanding, we use the longest token as the node label in the diagram, as it may contain the most information about the concept. When an annotated relationship describes a connection involving an entity that has not yet been streamed in from the response, we add a placeholder node indicating the incoming entity. Once the actual entity comes in, the placeholder node transforms into a real node (Figure~\ref{fig:ext}), allowing responsive construction of the diagram.

\subsection{Bidirectional Synchronization}
\label{sec:sync}

\system's diagram view serves as a rich and interactive interface for users to comprehend and explore information. However, users may occasionally need to refer back to the original textual response for details or to verify entities and relationships that may have been inaccurately extracted and visualized. To facilitate these processes, entity and relationship annotation information is stored and synchronized across different blocks of text and diagrams. When a user hovers over a node in the diagram, all co-referenced nodes in other individual diagrams are highlighted, as are the corresponding entities in the text and their originating sentences. This allows the user to quickly locate the term in the original response (Figure~\ref{fig:system}.5). Hovering over the edges in the diagram highlights the relationship tokens in the original response. Additionally, when a user selects a node, it highlights itself and the connecting edges in the diagram, which in turn highlights the corresponding set of tokens in the text response.

\begin{figure}[t]
    \centering
    \includegraphics[width=\linewidth]{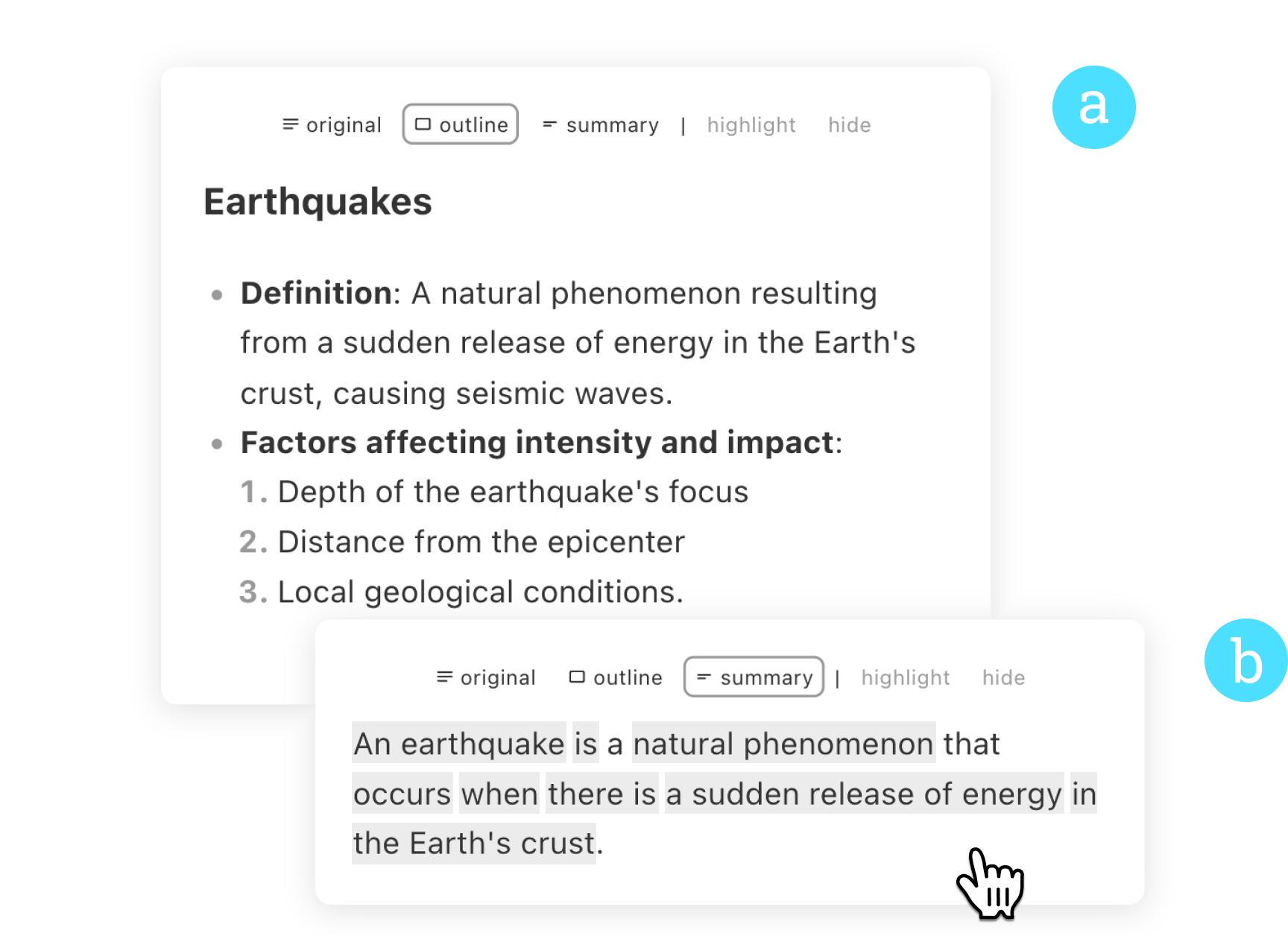}
    \caption{\rv{The Outline (a) and Summary (b) views.}}
    \Description{The Outline (a) and Summary (b) views.}
    \label{fig:views}
\end{figure}

\begin{figure}[t]
    \centering
    \includegraphics[width=\linewidth]{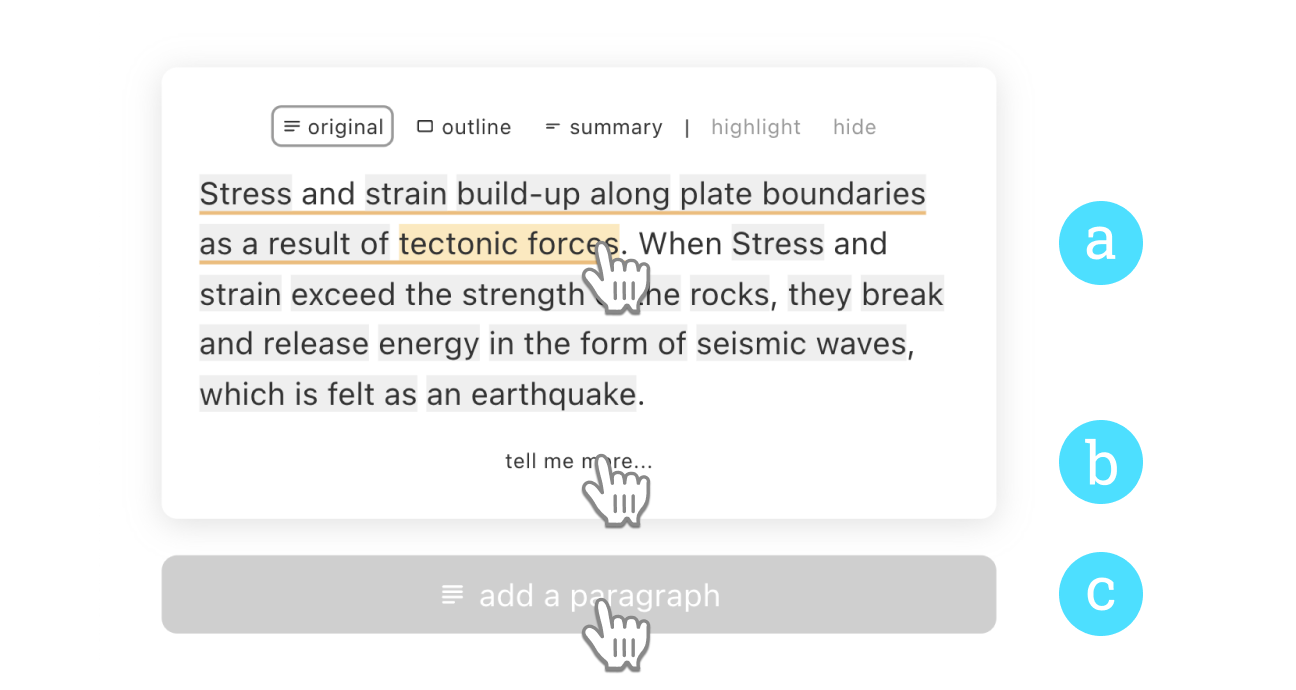}
    \caption{\rv{When a user hovers over an entity or relationship text token, it highlights itself, the co-references, and the corresponding nodes in the diagram (a). Hovering at the bottom of the text block reveals a ``tell me more'' button, which expands the paragraph (b). Clicking on ``add a paragraph'' button after the last text block prompts adding a new paragraph to the current response (c).}}
    \Description{When a user hovers over an entity and relationship text token, it highlights itself, the co-references, and the corresponding nodes in the diagram (a). Hovering at the bottom of the text block reveals a ``tell me more'' button, which expands the paragraph (b). Clicking on ``add a paragraph'' button after the last text block prompts adding a new paragraph to the current response (c).}
    \label{fig:hover}
\end{figure}

On the other hand, as users read the textual response to gain a detailed explanation of their topic of interest, they may switch to the diagram view from time to time to guide their comprehension of the long and unstructured text, for which they need to quickly locate the relevant node in the diagram. To support this, when the mouse cursor is hovering on the text, the corresponding nodes and relationships in the diagram are highlighted to enable quick navigation, as shown in Figure~\ref{fig:hover}.a. 

Users can collapse nodes to mitigate information overload as they progressively interact with the diagram (Section~\ref{sec:collapse}). When a node is collapsed, text tokens corresponding to the hidden leaf nodes are greyed out to reduce visual saliency compared to other parts of the response that correspond to active nodes in the diagram. This allows users to focus on important parts of the text and the diagram while minimizing distractions and clutter from less pertinent or uninterested information from both ends.

\subsection{Interaction with Diagram Nodes}

\system emphasizes using diagrams as the entry point and primary interface to gain information from \llms, supporting various information tasks to enhance user engagement and understanding, such as exploring unfamiliar concepts by accessing more detailed explanations and examples, and removing irrelevant content to focus on the most pertinent information.

\subsubsection{Continuous Exploration Through Explanation and Examples}

The initial response from \llms about an unfamiliar topic to the user could introduce more unfamiliar related concepts, for which they need to ask follow-up questions. Structuring the follow-up responses as diagrams and merging them in situ with the existing diagram allows contextual exploration and seamless integration of new information, which helps users understand the new concepts with the help of the existing diagrams. \system thus supports node-oriented exploration on top of the existing diagrams.

\begin{figure*}[ht]
    \centering
    \includegraphics[width=\textwidth]{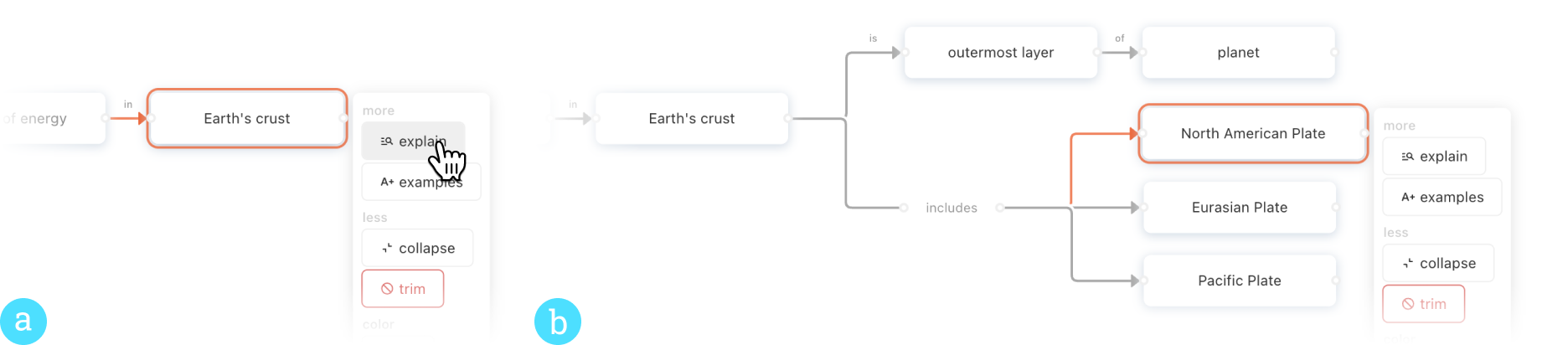}
    \caption{When the user is unfamiliar with the term `Earth's crust,' they select the corresponding node in the diagram and click the `Explain' button (a). This action generates a new segment of the diagram branching from the selected node, where the user can continue exploring examples and explanations of the new concepts (b).}
    \Description{When the user is unfamiliar with the term `Earth's crust,' they select the corresponding node in the diagram and click the `Explain' button (a). This action generates a new segment of the diagram branching from the selected node, where the user can continue exploring examples and explanations of the new concepts (b).}
    \label{fig:explain}
\end{figure*}

When a user encounters an unfamiliar concept in the diagram, they can select it and navigate to the \emph{Explain} or \emph{Examples} menu options to make a follow-up prompt for \gf to generate an explanation or a few examples about the concept (Figure~\ref{fig:explain}). The response text is directly appended to the end of the existing paragraph, and newly extracted entities and relationships are either co-referenced and assigned an existing identifier or cumulatively added to the diagram as new nodes and links.

\subsubsection{Reduced Information Overload Through Node Collapsing}
\label{sec:collapse}

In addition to gaining more information about a specific node, a user may want to remove the uninterested information or reduce the complexity of the diagrams. To achieve this, \system enables users to collapse and hide all leaf nodes of a diagram node, helping them concentrate on their areas of interest and proceed with their exploration in a less cluttered environment.

\subsubsection{User-Initiated Correction Through Trimming and Merging}
\label{sec:user-corr}

Even with the powerful GPT-4 and the complex prompts we use, incorrect labeling of entities and relationships still occurs. As the user explores, these errors accumulate and make the diagrams harder to follow and understand.
To address this, \gf supports the users to manually correct the errors if needed.

When a user finds a node is incorrectly extracted, e.g., a conjunctive adverb is identified as an entity, they can select the node and use the \emph{Trim} menu option to remove the entity from the diagram. As a result, relationships associating it with others and their inline annotations in the original response are also removed. When they identify two entities referring to the same concept but are not co-referenced correctly and are rendered as two nodes in the diagram, they can easily merge them by dragging one node on top of the other (to be merged as), after which the connecting relationships and their annotations will also be changed.

\begin{figure*}[ht]
    \centering
    \includegraphics[width=\textwidth]{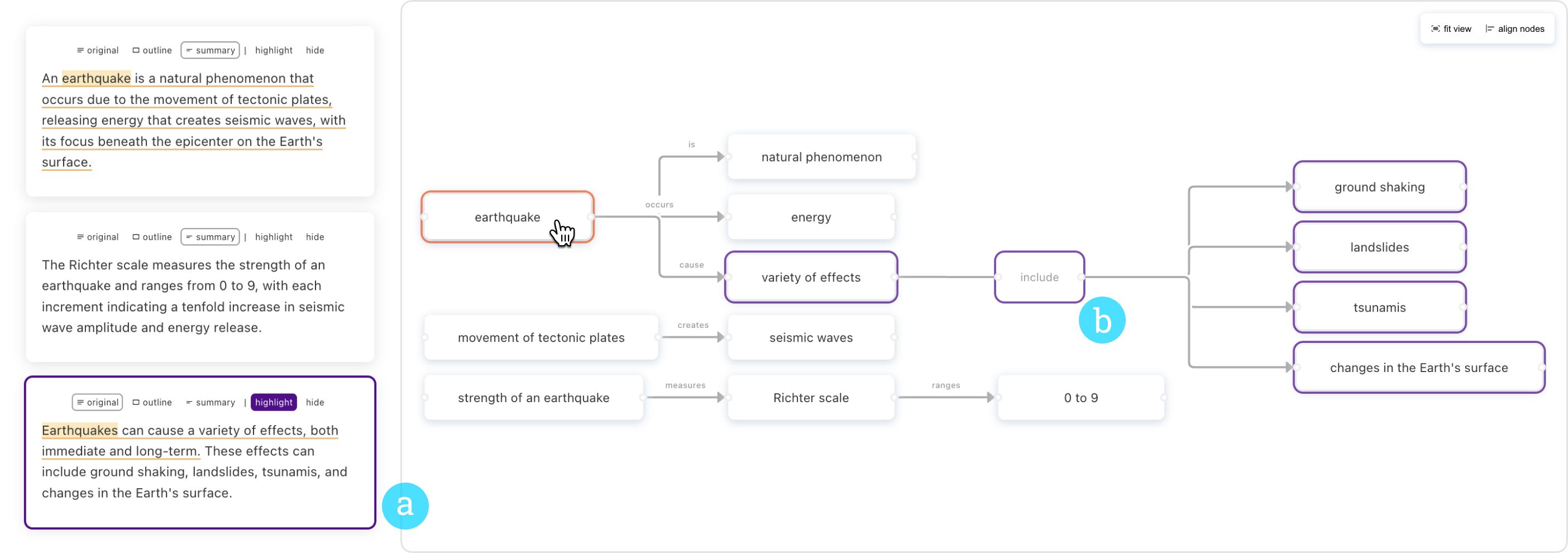}
    \caption{A merged diagram with a highlighted text block (a), whose entity nodes are highlighted in the diagram (b).}
    \Description{A merged diagram with a highlighted text block (a), whose entity nodes are highlighted in the diagram (b).}
    \label{fig:merged}
\end{figure*}

\subsection{Diagram-Level Information Managing}

While the diagram representation and a wide range of interactions supported by \system effectively help users view information from \llms in a structured and progressive way, the user can still be overwhelmed with a large amount of information and entity relationships contained in lengthy paragraphs. As introduced above, several diagram-level controls are introduced to help manage the information display complexity, including dividing responses into paragraphs, relationship saliency, and summary (Figure~\ref{fig:summary}.b). The text blocks can also be switched to the \emph{Outline} view that structures the plain text content with headings and bulleted lists (Figure~\ref{fig:views}.a). The underlining Markdown text is generated upon completion of each paragraph through prompt chaining, in parallel with self-correction for annotations and summarization (Appendix~\ref{app:p-corr}--\ref{app:p-outline}).

Conversely, if users find the provided information to be insufficient, they have the option to request additional details for each aspect (Figure~\ref{fig:hover}.b), or request the inclusion of more aspects related to the given topic (Figure~\ref{fig:hover}.c).

\subsection{Merging Diagrams}

Dividing \gf's raw response into different paragraphs and building a separate diagram for each of them effectively reduces information complexity and allows easy navigation and themed exploration. However, this structure is less helpful when users want to understand the relationships between aspects of the answer and get a holistic view of the topic, e.g., how a bird's lightweight body structure and respiratory system work cohesively to enable flight, or how Elon Musk's managing styles differ across his various companies. 

\system supports merging individual diagrams into one diagram for examination and integration (Figure~\ref{fig:merged}). The merging and splitting transition is animated to help users gain context for the individual and merged diagrams. When the user hovers over each of the text blocks, the corresponding nodes in the merged diagram get highlighted. If the user only wants to examine a subset of paragraphs, they can hide the others from the merged diagram, allowing any combination of diagrams to be merged and viewed. After the part gets hidden, \system automatically updates and animates the layout of the rest of the diagram to ensure an optimized and efficient view for the users.

\section{Technical Evaluation}

\begin{table*}[ht]
\centering
\caption{\rv{Technical Evaluation Results}\protect\footnotemark}
\label{tab:eva-table}
\begin{tabular}{ccllll}
\toprule
\multicolumn{1}{l}{} & \multicolumn{1}{l}{} & \textbf{\begin{tabular}[c]{@{}c@{}}Before\\ Correction\end{tabular}} & \textbf{\%} & \textbf{\begin{tabular}[c]{@{}c@{}}After One Round\\ of Correction\end{tabular}} & \textbf{\%} \\ \midrule
\multicolumn{1}{l}{} & Word Count & 4360 & \multicolumn{1}{l}{} & 4382 & \multicolumn{1}{l}{} \\ \hline\hline
\multirow{4}{*}{\textbf{\begin{tabular}[c]{@{}c@{}}Node\\ Annotation\end{tabular}}} & Total Entity Phrases & 1091 & \multicolumn{1}{l}{\multirow{4}{*}{}} & 1103 & \multicolumn{1}{l}{\multirow{4}{*}{}} \\
 & Total Extracted Entity Phrases & 1086 & \multicolumn{1}{l}{} & 1106 & \multicolumn{1}{l}{} \\
 & Correct Extracted Entity Phrases & 1043 & \multicolumn{1}{l}{} & 1074 & \multicolumn{1}{l}{} \\
 & Erroneous Extracted Entity Phrases & 43 & \multicolumn{1}{l}{} & 32 & \multicolumn{1}{l}{} \\ \hline
\multirow{3}{*}{Performance} & \textbf{Precision} & \multicolumn{1}{l}{\multirow{2}{*}{}} & \textbf{96.04\%} & \multicolumn{1}{l}{\multirow{2}{*}{}} & \textbf{97.11\%} \\
 & \textbf{Recall} & \multicolumn{1}{l}{} & \textbf{95.60\%} & \multicolumn{1}{l}{} & \textbf{97.37\%} \\
 & \textbf{F-score} & \multicolumn{1}{l}{} & \textbf{95.82\%} & \multicolumn{1}{l}{} & \textbf{97.24\%} \\ \hline
\multirow{4}{*}{Error Types} & Missing Entity Phrase & 48 & 4.40\% & 29 & 2.63\% \\
 & Incorrect Entity & 12 & 1.10\% & 13 & 1.18\% \\
 & Incomplete Entity & 22 & 2.02\% & 11 & 1.00\% \\
 & Incorrect Co-reference & 9 & 0.82\% & 8 & 0.73\% \\ \hline\hline
\multirow{4}{*}{\textbf{\begin{tabular}[c]{@{}c@{}}Relationship\\ Annotation\end{tabular}}} & Total Relationships & 825 & \multicolumn{1}{l}{\multirow{4}{*}{}} & 843 & \multicolumn{1}{l}{\multirow{4}{*}{}} \\
 & Total Extracted Relationships & 718 & \multicolumn{1}{l}{} & 813 & \multicolumn{1}{l}{} \\
 & Correct Extracted Relationships & 654 & \multicolumn{1}{l}{} & 765 & \multicolumn{1}{l}{} \\
 & Erroneous Extracted Relationships & 64 & \multicolumn{1}{l}{} & 48 & \multicolumn{1}{l}{} \\ \hline
\multirow{3}{*}{Performance} & \textbf{Precision} & \multicolumn{1}{l}{\multirow{2}{*}{}} & \textbf{91.09\%} & \multicolumn{1}{l}{\multirow{2}{*}{}} & \textbf{94.10\%} \\
 & \textbf{Recall} & \multicolumn{1}{l}{} & \textbf{79.27\%} & \multicolumn{1}{l}{} & \textbf{90.75\%} \\
 & \textbf{F-score} & \multicolumn{1}{l}{} & \textbf{84.77\%} & \multicolumn{1}{l}{} & \textbf{92.39\%} \\ \hline
\multirow{5}{*}{Error Types} & Missing Relationship & 171 & 20.73\% & 78 & 9.25\% \\
 & Dead-end Relationship & 37 & 4.48\% & 27 & 3.20\% \\
 & Reversed Relationship & 12 & 1.45\% & 10 & 1.19\% \\
 & Incomplete Relationship & 8 & 0.97\% & 7 & 0.83\% \\
 & Misattributed Relationship & 7 & 0.85\% & 4 & 0.47\% \\ \hline\hline
\multirow{2}{*}{\textbf{\begin{tabular}[c]{@{}c@{}}Detectable Errors\end{tabular}}} & \textbf{Orphan Nodes} & 133 & \textbf{12.25\%} & 38 & \textbf{3.44\%} \\
 & \textbf{Dead-end Relationships} & 37 & \textbf{4.48\%} & 27 & \textbf{3.20\%} \\ \bottomrule
\end{tabular}
\end{table*}

To gain a preliminary understanding of \gf's inline entity and relationship annotation ability, we conducted a small-scale technical evaluation to assess the accuracy of \gf inline annotations.

\subsection{Setup}
The goal of the evaluation is to assess the performance of our prompts with GPT-4 (model \code{gpt-4-0314}) in terms of initial inline entity and relationship annotations, as well as subsequent corrections. To achieve this, we simulated interactions with the system by utilizing GPT-4 to respond to a wide range of topics, following our prompting strategy. We then manually examined all the entity and relationship annotations in both the initial and self-corrected responses separately to gauge the system's performance.

\subsubsection{Topics} 
We developed a corpus with GPT-4 using a two-step method. Initially, we requested GPT-4 to generate a list of fifty areas encapsulating various facets of human knowledge (e.g., history, psychology, engineering, etc.). Subsequently, we prompted GPT-4 to provide responses on specific topics within each of these areas (e.g., The history of botanical gardens, Gamification). Following this, we amalgamated each topic with the \system{}’s prompt (Appendix~\ref{app:p-initial}) to collect the annotated responses.

\subsubsection{Error Coding} \rv{Three coders participated in the evaluation. Coders examined the annotations of each text response from GPT-4 and noted all incorrect annotations. An error taxonomy (Appendix~\mbox{\ref{app:ex}}) was iteratively established during the evaluation. While inter-coder reliability was not collected, each annotation was examined by two coders independently to minimize oversights.}

\rv{Besides syntactic errors, semantic errors were also identified. For example, Reversed Relationship errors indicate instances where the entity relationships were incorrectly inverted.} We found that, given a sentence, GPT-4 could produce different annotations at varied levels of granularity. For example, with this short phrase \mbox{\textit{``the goal of the evaluation,''}} GPT-4 could produce two different annotations, such as \code{[the goal (\$N1)] [of (\$L, \$N1, \$N2)] [the evaluation (\$N2)]} or \code{[the goal of the evaluation (\$N1)]}. When examining the annotations, as long as they were semantically correct, regardless of the granularity, we accepted them as correct annotations. \footnotetext{Detailed explanation and examples of the error types can be found in Appendix~\ref{app:ex}.}

\subsubsection{Initial and Self-Corrected Responses} In response to all 50 queries, GPT-4 cumulatively returned initial responses encompassing 4360 words, 1086 annotated entity phrases, and 718 annotated relationships. Errors that appear in the initial responses were aggregated and listed in column \emph{Before Correction} in Table~\ref{tab:eva-table}.

\system{} is capable of detecting Orphan Nodes (i.e., entities without any associated relationships, usually resulting from Incorrect Entities or Incomplete Relationships) or Dead-end Relationships (i.e., relationships attempting to connect non-existent entities, typically due to Dead-end Relationships), as listed in rows \emph{Detectable Errors} (errors that can be detected by \system{}) in Table~\ref{tab:eva-table}. When errors are detected, \gf can conduct a round of corrections with GPT-4 using a correction prompt (Appendix~\ref{app:p-corr}).

We combined the corrected responses with initial responses that didn't include these errors. The total dataset contained 4382 words (the system is instructed to rewrite the sentence if necessary), 1106 annotated entity phrases, and 813 annotated relationships. Errors that emerged in these responses were consolidated and presented in column \emph{After One Round of Correction} in Table~\ref{tab:eva-table}).

\subsection{Key Findings}

We consolidate our primary observations pertaining to the initial annotating performance and the improvements observed after a single round of correction.

\subsubsection{Initial Annotation Performance (Before Correction)}

For entity annotation, we observe an F-score of 95.82\%. For relationship annotation, we note an F-score of 84.77\%. The majority of errors in entity annotation arise due to Missing Entity Phrases ($n=48$) and Incomplete Entities ($n=22$). Conversely, the bulk of errors in relationship annotations results from Missing Relationships ($n=171$) and Dead-end Relationships ($n=37$).
The annotated responses include 133 Orphan Nodes (attributable to Incorrect Entities or Misattributed Relationships) and 37 Dead-end Relationships.

\subsubsection{Annotation Performance with One Round of Correction}

Upon correction, the precision for entity annotation improves to 97.11\% (increases 1.07\%), and the recall rises to 97.37\% (increases 1.77\%), resulting in an F-score of 97.24\%. For relationship annotations, we notice a precision of 94.10\% (increases 3.01\%) and a recall of 90.75\% (increases 11.48\%), resulting in an F-score of 92.39\%. The numbers of Orphan Nodes and Dead-end Relationships drop to 38 and 27, respectively. Most errors in entity annotations stem from Missing and Incorrect Entities. Similarly, for relationship annotations, most errors originate from Missing and Dead-end Relationships.

\subsection{Summary}

In summary, the technical evaluation results suggest that the annotation and correction prompts employed by \system{} reliably execute entity and relationship annotation tasks with our prompting and self-correction strategies. Specifically, they achieve an F-score of 97.24\% for entity annotation, and an F-score of 92.39\% for relationship annotation, with one round of correction.

\section{User Evaluation}
\label{sec:conclusion}

The primary goal of this project is to investigate the potential benefits of leveraging graphical representations for textual information from LLMs. This approach aims to address many of their limitations and enhance user interactions, especially during exploratory information-seeking tasks. In order to assess the efficacy of this novel approach, we conducted a user evaluation.

\subsection{Participants}

We recruited seven experienced users of ChatGPT to evaluate \system. To ensure their level of experience, we requested participants to share screenshots of their ChatGPT history, enabling us to gauge the complexity of the prompts they typically used. Each participant was compensated with 30 USD for one hour of their participation.

\subsection{Setup}
We conducted all sessions remotely using Zoom, with \system deployed on a cloud server for participants to access. We recorded the audio and screen activity during each session, subsequently transcribing these records to facilitate inductive qualitative analysis using the thematic analysis method~\cite{braun2012thematic}.

\subsection{Procedure}

A study session consisted of four steps.

\subsubsection{Introduction (5 minutes)}
The interviewer briefly introduced the goal of the project and study and presented four potential topics for exploration: Neuro-divergence, Supply and Demand, Northern Lights, and Inflation. The participant was then asked to select one of these subjects of interest.

\subsubsection{System Training (10 minutes)}
The interviewer proceeded with a guided tour of all key features of \system{}, with a standard example topic---Electric Vehicles. This step aimed to familiarize participants with \system{} interface and functionalities.

\subsubsection{System Interaction (20 minutes)}
Participants were guided to use \system{} in the completion of their tasks. Specifically, they were tasked to gather information to prepare for a hypothetical lecture. They were given the following prompt: \textit{``Imagine you are a professor at a public university, and you are scheduled to deliver a lecture on the topic of [selected topic]. Here is a list of concepts (three related concepts and one question were provided for each topic) that you aim to explain to your students by the end of the lecture.''}

\subsubsection{Survey and Interview (20 minutes)}
Following the completion of the tasks, participants filled out a survey that consisted of questions designed to gauge their perceptions of the system's usefulness and the utility of its multiple representations of information.
Each question was rated on a 5-point Likert scale (1---strongly disagree, and 5---strongly agree).
Later, we conducted an interview wherein participants were prompted to draw comparisons between their current experience with \system and their prior interactions with ChatGPT in the context of knowledge acquisition tasks. The purpose was to identify how our system addresses the challenges that typically arise when using ChatGPT.

\subsection{Results}

We share the results of our user evaluation, detailing the effectiveness of \system{} in facilitating the understanding of information, managing the complexity of diagrams, and the distinct workflows when interacting with typical LLMs and \system{}. We also discuss the identified limitations of the current system.

\begin{figure}[ht]
    \centering
    \includegraphics[width=\linewidth]{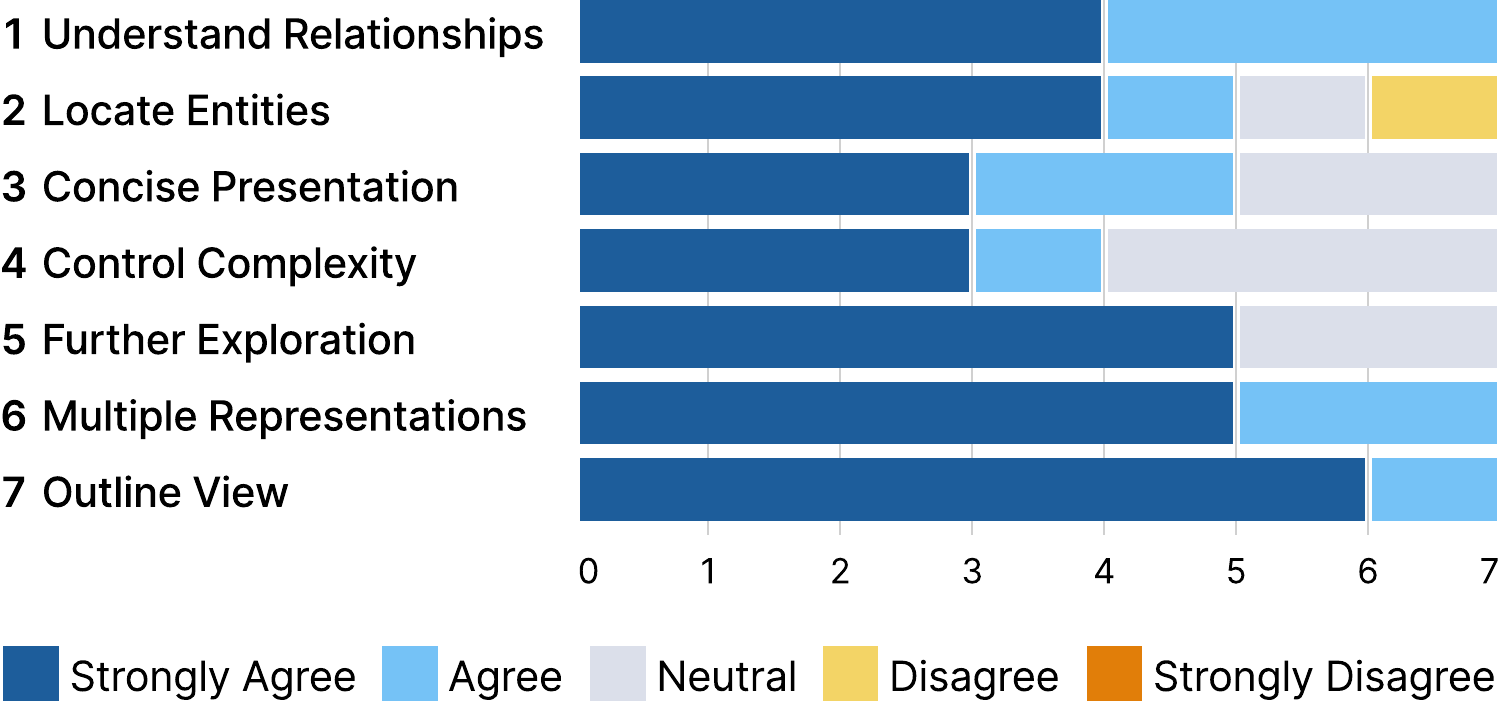}
    \caption{\rv{Participants' responses to utility and usability of \mbox{\system} interface and various features, measured on a 5-point Likert scale.}}
    \Description{Participants' responses to utility and usability of \system interface and various features, measured on a 5-point Likert scale.}
    \label{fig:likert}
\end{figure}

\subsubsection{\system{} Facilitates Information Comprehension}
Participants found that the node-link diagrams enhanced their understanding of the diverse relationships inherent in the topic they explored (Figure~\ref{fig:likert}.1). For instance, P5 stated that diagrams helped \textit{``visualize the connections,''} and P4 suggested they aid in comprehending how different aspects connect. Participants particularly praised the diagrams for providing an \textit{``overall view of a topic''} and for \textit{``understanding a set of instructions''} (P8), likening it to a \textit{``mind map''} (P8), and stating that it provides an understanding of the organization of information (P5). Conversely, they noted that conventional text responses from LLMs can be \textit{``wordy,''} while diagrams make it \textit{``faster to get the information''} (P1).

The bidirectional mapping between the diagram and paragraph through highlighting \textit{``is a good visual cue to get the users' attention to understand the various relationships''}~(P5) and helps users locate and comprehend various terms and concepts from the diagram with explanations easily (Figure~\ref{fig:likert}.2).

\subsubsection{\system{} Provides Sufficient Control for Diagram Complexity}
Most participants found the amount of information presented in the responses to be concise (Figure~\ref{fig:likert}.3), and they appreciated the ability to control the level of detail they wished to see with \system{} (Figure~\ref{fig:likert}.4). For instance, the presentation of smaller diagrams for each individual paragraph was praised for helping understand \textit{``what (was) happening within a single paragraph''} (P7). P1 expressed satisfaction with the degree of control they had over the depth and breadth of information within each paragraph.

Participants responded favorably to the ability to split and merge diagrams, recognizing that this functionality effectively supports diverse information consumption goals. P1 highlighted the helpfulness of split diagrams when they \textit{``care about one particular paragraph,''} while appreciating the merged diagram when they need to \textit{``get the overall information of a particular topic.''} P2 suggested that the capacity to flexibly switch between these views could assist students in self-learning tasks, maintaining focus on both critical details and the broader picture. P5 highlighted that merged view offers insights beyond \textit{``standalone concepts.''}
For example, when exploring neuro-diversity, P7 used split diagrams to understand lower-level concepts in detail, such as strategies to identify neuro-divergence in individuals. Then, they used the merged diagram to get an overview of how different strategies for identification can be integrated. The ability to highlight parts enabled them to examine how individual sections contributed to and were integrated within the larger diagram (P1, P2, P5).

\subsubsection{\system{} Reduces Prompting Effort and Facilitates Exploration}
While using ChatGPT, which often generates verbose information, the burden is typically on the user to craft prompts that control the length of the responses, for instance, `use less than 200 words.' With \system{}, however, participants found the extensive controls for managing complexity effectively reduced the need for such directive prompts. As P1 noted, \textit{``there's no need to [specify] `make it brief,' cause you can just click on [the interface].''}

Participants noted how \system{} made it convenient to extract new information through interactions with the diagrams (Figure~\ref{fig:likert}.5). \system made it particularly easy to construct prompts for creating examples and explanations during learning activities, which can often be monotonous and time-consuming. As P1 noted, with ChatGPT, users have to prompt \textit{``Please incorporate more examples about `this thing' in your answer.''} However, with \system{}, acquiring context-specific examples is straightforward: \textit{``clicking the `Examples' button just does it, and (you) don't have to think of another prompt''} (P5). Extending this point, P8 stated that following a \textit{``chain''} becomes simple without the need to write a prompt. P4, for example, intended to understand the meaning of `particles' in \textit{``charged particles from the sun.''} They selected the node `particles' in the diagram and clicked the `Explain' button. This action extended the paragraph and constructed a new part of the diagram stemming from the node, serving to clarify the context-specific meaning of `particles.' P4 proceeded to follow this process for many other terms that emerged in the explanation, such as protons and atomic nuclei.

\subsubsection{\system{} Combines the Strengths of Multiple Representations}

While the diagrams serve as the primary representation of information, \system{} also provides the original text, an outline, and a summary for each paragraph. Participants found these alternative representations to have complementary strengths. For instance, the outline view, formatted with headings and lists, effectively organized and highlighted the key ideas in each paragraph (Figure~\ref{fig:likert}.7), like a \textit{``lecture slide''} that significantly aids learning (P2). Conversely, the diagrams offset the limitations of the outline view, which simply enumerates points without showing their interconnections. The diagrams, however, clearly illustrate the interactions among the bulleted points and \textit{``how they relate back to the main idea''} (P4). P8 explored the concept of `marginal utility' and found themselves trying to synthesize a variety of encompassed economics jargon such as `total utility' and `welfare programs.'
They switched the text block to the outline view, which, coupled with the diagram, provided hierarchical and relational information detailing how these terms function within the concept of `marginal utility.'

Participants also found that the original text could fill in the details missing from the diagrams. For instance, P1 enjoyed highlighting specific parts of the text they were interested in to get more details, which then formed the basis when seeking additional information. P5 reported that at times, it could be \textit{``a little intimidating if you just see the whole diagram,''} and the synchronization between the outline view and the diagram allowed them to use the outline view as a way to filter out unnecessary nodes in the diagrams. These examples not only illustrate the strengths and weaknesses of each representation, but also show that participants could adaptively employ, switch, and combine them for different use cases.

\subsubsection{Limitations}

\system{} separates the response into smaller paragraphs to reduce the complexity of diagrams. However, this can lead to extensive use of screen space, and P5 suggested that they need to do \textit{``a lot of scrolling''} to view all the diagrams. While more complex diagrams can be space-efficient, they might compromise understanding. To address this inherent trade-off between diagram complexity and spatial efficiency, in addition to merging all small diagrams into one, future iterations of the system need to provide more control for users to flexibly merge any selection of diagrams.

Some participants reported difficulties in understanding the diagrams when they did not match their mental models or included an overwhelming amount of details. On the other hand, while text responses from \gf tend to be verbose, the text representation does not immediately impose a mental structure and allows users to gradually construct their own mental models while reading the text, unlike diagrams. \system{} alleviates this problem by providing both the generated diagrams and the original text. This issue can be further mitigated by enabling users to manipulate the diagrams during generation to align the diagrams with their mental models.

Other limitations associated with the current interface design, annotation performance, and user expertise were also identified. P1 found the animations used to progressively augment diagrams distracting. Some participants found the generation latency, due to GPT-4's performance, negatively impacted the experience. These limitations can be addressed by employing smoother animation effects, prompt engineering to minimize annotation errors, and more responsive generative models. \rv{On the other hand, as participants were allowed to choose the topic of their preference, they invariably chose ones they were either `Familiar' or `Very Familiar' with. Future studies could focus on assessing how the system might assist users of varying familiarity levels with a given topic.}

\subsubsection{Summary}
The study findings demonstrate that the prompting techniques and the interface designs leveraged by \system{} offer a more direct representation of the concepts and their relationships, enhancing the comprehension of information from LLMs. The rich and flexible control over the complexity of the diagrams, along with the combination of the strengths of various representations provided by \system{}, enabled participants to maintain control over the amount of information they wished to consume for diverse information-seeking goals and tasks.

\section{Discussion}
\label{sec:evaluation}
Our user study findings highlight the benefits, limitations, and opportunities of \system{}, which employs graphical representation and enhanced interaction with LLM-generated information.
We discuss these aspects in detail below.

\subsection{Improving Annotation Performance}
While our technical evaluation demonstrates the advanced inline annotation capabilities of GPT-4, especially with self-correction, the final annotations still contain errors, such as relationships that point to non-existent entities. These annotation errors can produce misleading diagrams, for which the users need to cross-reference the original text to alleviate their confusion.

\rv{Many approaches can be utilized to improve annotation performance. For example, incorporating domain knowledge as references for corresponding annotation tasks. Improved prompt designs informed by new knowledge of \mbox{\llms} could also help improve the overall performance, e.g., assigning roles to the model for different domain-specific tasks or providing more training examples. A comparison with the traditional NLP methods in Named Entity Recognition (NER) and Semantic Role Labeling (SRL) could reveal potential LLM-specific biases in annotation, which could also guide us in refining our prompting strategies~\mbox{\cite{palmer2011semantic,marquez2008semantic,tareq2020named}}.
On the other hand, a more advanced and fine-tuned model could also lead to improved annotation performance.}

\subsection{Prompting \llms via Graphical Interfaces}
 
The graphical user interface of \system enables users to employ direct manipulation with the diagram to request explanations and examples from \llms, saving users' efforts to manually craft textual prompts. Future work could provide more options for interacting with \llms graphically. For example, the system can allow users to select multiple disconnected nodes and build a new diagram illustrating their connections, or summarize a branch of a diagram with one higher-level concept. Moreover, a text input box can be provided to allow customized requests. As the user explores the knowledge space through the graphical interface, their prior actions and the current diagram can be leveraged as context to construct prompts for \llms to get responses that are better aligned with the user's needs. For instance, when a user collapses a branch, the subsequent prompts can incorporate text that indicates the collapsed aspect is less relevant to the user's present goal.

\subsection{Supporting Applications Across Diverse Domains}
The node-link diagrams constructed by \system, utilizing \llm responses, are particularly helpful for tasks requiring a comprehensive understanding of diverse concepts and the interconnections among them, such as exploratory information seeking~\cite{palani2022interweave}. Participants from our user study identified several immediate applications of \system, including education and professional training. Future research could explore the application of \system's \emph{real-time diagram generation capabilities} to other contexts where node-link diagrams serve as a powerful visual facilitator.
% For instance, these diagrams could be used to visualize character relationships in fictional writing or to depict connections among academic papers in literature.
For instance, in the context of group discussion, diagrams can be generated based on conversations to visualize the discussed concepts and their connections, acting as a collective knowledge map to ground and facilitate the discussions.

\rv{Information accuracy is critical to many applications, such as academic literature reviews. However, current LLMs are prone to ``hallucination'' and can generate factually incorrect text~\mbox{\cite{zhang2023language,azamfirei2023large}}. Diagrams could help verify LLM-generated information by breaking a large body of text into pieces that can be individually validated. For example, a pair of connected nodes and their relationships could be treated as a unit piece of information that can be validated against external knowledge bases, and relationships of different degrees of certainty can be visualized with different color intensities.}

\subsection{Exploring Representations Beyond Node-Link Diagrams}
\rv{A theme that surfaced in the user study is that although node-link diagrams can serve as a suitable representation for understanding interconnected concepts and relationships within LLM responses, they might not always be optimal for a wide range of information tasks~\mbox{\cite{r40}}. Other formats, such as tables, storyboards, animations, and flowcharts, can be more suitable representations for different aspects of the information. For instance, a table can be clearer when comparing different aspects of multiple concepts, and animations can better illustrate dynamic processes~\mbox{\cite{TVERSKY2002247}}. Future research could investigate how to annotate and construct such representation formats and intelligently select the most suitable one according to the context. Other systems could also explore creating connections among these representations and offering ways for users to switch between them as needed and desired.}

\section{Conclusion}
\llms such as \gf have swiftly gained recognition and popularity due to their unprecedented intelligence and potential for a wide range of applications. Existing interfaces for \llms, like ChatGPT, employ linear and text-based interfaces, often generating an abundance of information. Our formative study identified three challenges related to the limited usability, readability, and interactivity of textual \llm responses. \system, in contrast, leverages dynamic and interactive diagrams as the primary interface for interacting with \llms. Our user study indicated that \system effectively addressed many limitations inherent in conversational interfaces by offering flexible graphical representations that facilitated direct and adaptable graphical dialogue with LLMs.

%%
% Acknowledgments.
\begin{acks}
This work would not be possible without the selfless and heartwarming support of all the members of the Creativity Lab at UC San Diego. Our deepest gratitude extends to Fuling Sun, who was crucial in facilitating a productive drive toward the successful completion of this project. We would also like to thank Sangho Suh, Bryan Min, Matthew Beaudouin-Lafon, and Jane E for helping with the video figure production, and William Duan, Vidya Madhavan, Tony Meng, Xiaoshuo Yao, and Juliet (Lingye) Zhuang for assistance with the technical evaluation, as well as Brian Hempel and Devamardeep Hayatpur for proofreading the paper draft. We thank anonymous reviewers for their constructive and insightful reviews. NSF grant \#2009003 provided financial support.
\end{acks}

\bibliographystyle{ACM-Reference-Format}
\bibliography{uist23-25}


\begin{thebibliography}{94}

%%% the Web address.

\ifx \showCODEN    \undefined \def \showCODEN     #1{\unskip}     \fi
\ifx \showDOI      \undefined \def \showDOI       #1{#1}\fi
\ifx \showISBNx    \undefined \def \showISBNx     #1{\unskip}     \fi
\ifx \showISBNxiii \undefined \def \showISBNxiii  #1{\unskip}     \fi
\ifx \showISSN     \undefined \def \showISSN      #1{\unskip}     \fi
\ifx \showLCCN     \undefined \def \showLCCN      #1{\unskip}     \fi
\ifx \shownote     \undefined \def \shownote      #1{#1}          \fi
\ifx \showarticletitle \undefined \def \showarticletitle #1{#1}   \fi
\ifx \showURL      \undefined \def \showURL       {\relax}        \fi
% invisible to TeX
\providecommand\bibfield[2]{#2}
\providecommand\bibinfo[2]{#2}
\providecommand\natexlab[1]{#1}
\providecommand\showeprint[2][]{arXiv:#2}

\bibitem[Agrawala et~al\mbox{.}(2011)]%
        {agrawala2011design}
\bibfield{author}{\bibinfo{person}{Maneesh Agrawala}, \bibinfo{person}{Wilmot
  Li}, {and} \bibinfo{person}{Floraine Berthouzoz}.}
  \bibinfo{year}{2011}\natexlab{}.
\newblock \showarticletitle{Design principles for visual communication}.
\newblock \bibinfo{journal}{\emph{Commun. ACM}} \bibinfo{volume}{54},
  \bibinfo{number}{4} (\bibinfo{year}{2011}), \bibinfo{pages}{60--69}.
\newblock


\bibitem[Ainsworth(2006)]%
        {AINSWORTH2006183}
\bibfield{author}{\bibinfo{person}{Shaaron Ainsworth}.}
  \bibinfo{year}{2006}\natexlab{}.
\newblock \showarticletitle{DeFT: A conceptual framework for considering
  learning with multiple representations}.
\newblock \bibinfo{journal}{\emph{Learning and Instruction}}
  \bibinfo{volume}{16}, \bibinfo{number}{3} (\bibinfo{year}{2006}),
  \bibinfo{pages}{183--198}.
\newblock
\showISSN{0959-4752}
\urldef\tempurl%
\url{https://doi.org/10.1016/j.learninstruc.2006.03.001}
\showDOI{\tempurl}


\bibitem[Ainsworth and Th~Loizou(2003)]%
        {ainsworth2003effects}
\bibfield{author}{\bibinfo{person}{Shaaron Ainsworth} {and}
  \bibinfo{person}{Andrea Th~Loizou}.} \bibinfo{year}{2003}\natexlab{}.
\newblock \showarticletitle{The effects of self-explaining when learning with
  text or diagrams}.
\newblock \bibinfo{journal}{\emph{Cognitive science}} \bibinfo{volume}{27},
  \bibinfo{number}{4} (\bibinfo{year}{2003}), \bibinfo{pages}{669--681}.
\newblock


\bibitem[Al-Moslmi et~al\mbox{.}(2020)]%
        {tareq2020named}
\bibfield{author}{\bibinfo{person}{Tareq Al-Moslmi}, \bibinfo{person}{Marc
  Gallofré~Ocaña}, \bibinfo{person}{Andreas L.~Opdahl}, {and}
  \bibinfo{person}{Csaba Veres}.} \bibinfo{year}{2020}\natexlab{}.
\newblock \showarticletitle{Named Entity Extraction for Knowledge Graphs: A
  Literature Overview}.
\newblock \bibinfo{journal}{\emph{IEEE Access}}  \bibinfo{volume}{8}
  (\bibinfo{year}{2020}), \bibinfo{pages}{32862--32881}.
\newblock
\urldef\tempurl%
\url{https://doi.org/10.1109/ACCESS.2020.2973928}
\showDOI{\tempurl}


\bibitem[Azamfirei et~al\mbox{.}(2023)]%
        {azamfirei2023large}
\bibfield{author}{\bibinfo{person}{Razvan Azamfirei}, \bibinfo{person}{Sapna~R
  Kudchadkar}, {and} \bibinfo{person}{James Fackler}.}
  \bibinfo{year}{2023}\natexlab{}.
\newblock \showarticletitle{Large language models and the perils of their
  hallucinations}.
\newblock \bibinfo{journal}{\emph{Critical Care}} \bibinfo{volume}{27},
  \bibinfo{number}{1} (\bibinfo{year}{2023}), \bibinfo{pages}{1--2}.
\newblock


\bibitem[Badam et~al\mbox{.}(2018)]%
        {elasticdocs2018}
\bibfield{author}{\bibinfo{person}{Sriram~Karthik Badam},
  \bibinfo{person}{Zhicheng Liu}, {and} \bibinfo{person}{Niklas Elmqvist}.}
  \bibinfo{year}{2018}\natexlab{}.
\newblock \showarticletitle{Elastic documents: Coupling text and tables through
  contextual visualizations for enhanced document reading}.
\newblock \bibinfo{journal}{\emph{IEEE transactions on visualization and
  computer graphics}} \bibinfo{volume}{25}, \bibinfo{number}{1}
  (\bibinfo{year}{2018}), \bibinfo{pages}{661--671}.
\newblock


\bibitem[Baidoo-Anu and Owusu~Ansah(2023)]%
        {baidoo2023education}
\bibfield{author}{\bibinfo{person}{David Baidoo-Anu} {and}
  \bibinfo{person}{Leticia Owusu~Ansah}.} \bibinfo{year}{2023}\natexlab{}.
\newblock \showarticletitle{Education in the era of generative artificial
  intelligence (AI): Understanding the potential benefits of ChatGPT in
  promoting teaching and learning}.
\newblock \bibinfo{journal}{\emph{Available at SSRN 4337484}}
  (\bibinfo{year}{2023}).
\newblock
\urldef\tempurl%
\url{http://dx.doi.org/10.2139/ssrn.4337484}
\showURL{%
\tempurl}


\bibitem[Baker et~al\mbox{.}(2009)]%
        {baker2009using}
\bibfield{author}{\bibinfo{person}{Jeff Baker}, \bibinfo{person}{Donald Jones},
  {and} \bibinfo{person}{Jim Burkman}.} \bibinfo{year}{2009}\natexlab{}.
\newblock \showarticletitle{Using visual representations of data to enhance
  sensemaking in data exploration tasks}.
\newblock \bibinfo{journal}{\emph{Journal of the Association for Information
  Systems}} \bibinfo{volume}{10}, \bibinfo{number}{7} (\bibinfo{year}{2009}),
  \bibinfo{pages}{2}.
\newblock


\bibitem[Bederson and Hollan(1994)]%
        {pad++1994}
\bibfield{author}{\bibinfo{person}{Benjamin~B. Bederson} {and}
  \bibinfo{person}{James~D. Hollan}.} \bibinfo{year}{1994}\natexlab{}.
\newblock \showarticletitle{Pad++: A Zooming Graphical Interface for Exploring
  Alternate Interface Physics}. In \bibinfo{booktitle}{\emph{Proceedings of the
  7th Annual ACM Symposium on User Interface Software and Technology}} (Marina
  del Rey, California, USA) \emph{(\bibinfo{series}{UIST '94})}.
  \bibinfo{publisher}{Association for Computing Machinery},
  \bibinfo{address}{New York, NY, USA}, \bibinfo{pages}{17–26}.
\newblock
\showISBNx{0897916573}
\urldef\tempurl%
\url{https://doi.org/10.1145/192426.192435}
\showDOI{\tempurl}


\bibitem[Bolt(1980)]%
        {bolt1980put}
\bibfield{author}{\bibinfo{person}{Richard~A. Bolt}.}
  \bibinfo{year}{1980}\natexlab{}.
\newblock \showarticletitle{``Put-That-There'': Voice and Gesture at the
  Graphics Interface}. In \bibinfo{booktitle}{\emph{Proceedings of the 7th
  Annual Conference on Computer Graphics and Interactive Techniques}} (Seattle,
  Washington, USA) \emph{(\bibinfo{series}{SIGGRAPH '80})}.
  \bibinfo{publisher}{Association for Computing Machinery},
  \bibinfo{address}{New York, NY, USA}, \bibinfo{pages}{262–270}.
\newblock
\showISBNx{0897910214}
\urldef\tempurl%
\url{https://doi.org/10.1145/800250.807503}
\showDOI{\tempurl}


\bibitem[Borsos et~al\mbox{.}(2022)]%
        {borsos2022audiolm}
\bibfield{author}{\bibinfo{person}{Zal\'{a}n Borsos},
  \bibinfo{person}{Rapha\"{e}l Marinier}, \bibinfo{person}{Damien Vincent},
  \bibinfo{person}{Eugene Kharitonov}, \bibinfo{person}{Olivier Pietquin},
  \bibinfo{person}{Matt Sharifi}, \bibinfo{person}{Olivier Teboul},
  \bibinfo{person}{David Grangier}, \bibinfo{person}{Marco Tagliasacchi}, {and}
  \bibinfo{person}{Neil Zeghidour}.} \bibinfo{year}{2022}\natexlab{}.
\newblock \bibinfo{title}{AudioLM: a Language Modeling Approach to Audio
  Generation}.
\newblock
\newblock
\showeprint[arxiv]{2209.03143}~[cs.SD]


\bibitem[Braun and Clarke(2012)]%
        {braun2012thematic}
\bibfield{author}{\bibinfo{person}{Virginia Braun} {and}
  \bibinfo{person}{Victoria Clarke}.} \bibinfo{year}{2012}\natexlab{}.
\newblock \bibinfo{booktitle}{\emph{Thematic analysis.}}
\newblock \bibinfo{publisher}{American Psychological Association}.
\newblock


\bibitem[Brusilovsky(2001)]%
        {brusilovsky2001adaptive}
\bibfield{author}{\bibinfo{person}{Peter Brusilovsky}.}
  \bibinfo{year}{2001}\natexlab{}.
\newblock \showarticletitle{Adaptive hypermedia}.
\newblock \bibinfo{journal}{\emph{User modeling and user-adapted interaction}}
  \bibinfo{volume}{11} (\bibinfo{year}{2001}), \bibinfo{pages}{87--110}.
\newblock


\bibitem[Bubeck et~al\mbox{.}(2023)]%
        {sparks2023}
\bibfield{author}{\bibinfo{person}{Sébastien Bubeck}, \bibinfo{person}{Varun
  Chandrasekaran}, \bibinfo{person}{Ronen Eldan}, \bibinfo{person}{Johannes
  Gehrke}, \bibinfo{person}{Eric Horvitz}, \bibinfo{person}{Ece Kamar},
  \bibinfo{person}{Peter Lee}, \bibinfo{person}{Yin~Tat Lee},
  \bibinfo{person}{Yuanzhi Li}, \bibinfo{person}{Scott Lundberg},
  \bibinfo{person}{Harsha Nori}, \bibinfo{person}{Hamid Palangi},
  \bibinfo{person}{Marco~Tulio Ribeiro}, {and} \bibinfo{person}{Yi Zhang}.}
  \bibinfo{year}{2023}\natexlab{}.
\newblock \bibinfo{title}{Sparks of Artificial General Intelligence: Early
  experiments with GPT-4}.
\newblock
\newblock
\showeprint[arxiv]{2303.12712}~[cs.CL]


\bibitem[Ca{\~{n}}as et~al\mbox{.}(2005)]%
        {Conceptmaps2005}
\bibfield{author}{\bibinfo{person}{Alberto~J. Ca{\~{n}}as},
  \bibinfo{person}{Roger Carff}, \bibinfo{person}{Greg Hill},
  \bibinfo{person}{Marco Carvalho}, \bibinfo{person}{Marco Arguedas},
  \bibinfo{person}{Thomas~C. Eskridge}, \bibinfo{person}{James Lott}, {and}
  \bibinfo{person}{Rodrigo Carvajal}.} \bibinfo{year}{2005}\natexlab{}.
\newblock \bibinfo{booktitle}{\emph{Concept Maps: Integrating Knowledge and
  Information Visualization}}.
\newblock \bibinfo{publisher}{Springer Berlin Heidelberg},
  \bibinfo{address}{Berlin, Heidelberg}, \bibinfo{pages}{205--219}.
\newblock
\showISBNx{978-3-540-31962-7}
\urldef\tempurl%
\url{https://doi.org/10.1007/11510154_11}
\showDOI{\tempurl}


\bibitem[Card et~al\mbox{.}(1999)]%
        {card1999}
\bibfield{editor}{\bibinfo{person}{Stuart~K. Card}, \bibinfo{person}{Jock~D.
  Mackinlay}, {and} \bibinfo{person}{Ben Shneiderman}} (Eds.).
  \bibinfo{year}{1999}\natexlab{}.
\newblock \bibinfo{booktitle}{\emph{Readings in Information Visualization:
  Using Vision to Think}}.
\newblock \bibinfo{publisher}{Morgan Kaufmann Publishers Inc.},
  \bibinfo{address}{San Francisco, CA, USA}.
\newblock
\showISBNx{1558605339}


\bibitem[Chang et~al\mbox{.}(2015)]%
        {chang2015text}
\bibfield{author}{\bibinfo{person}{Angel Chang}, \bibinfo{person}{Will Monroe},
  \bibinfo{person}{Manolis Savva}, \bibinfo{person}{Christopher Potts}, {and}
  \bibinfo{person}{Christopher~D. Manning}.} \bibinfo{year}{2015}\natexlab{}.
\newblock \bibinfo{title}{Text to 3D Scene Generation with Rich Lexical
  Grounding}.
\newblock
\newblock
\showeprint[arxiv]{1505.06289}~[cs.CL]


\bibitem[Chang et~al\mbox{.}(2023)]%
        {chang2023muse}
\bibfield{author}{\bibinfo{person}{Huiwen Chang}, \bibinfo{person}{Han Zhang},
  \bibinfo{person}{Jarred Barber}, \bibinfo{person}{AJ Maschinot},
  \bibinfo{person}{Jose Lezama}, \bibinfo{person}{Lu Jiang},
  \bibinfo{person}{Ming-Hsuan Yang}, \bibinfo{person}{Kevin Murphy},
  \bibinfo{person}{William~T. Freeman}, \bibinfo{person}{Michael Rubinstein},
  \bibinfo{person}{Yuanzhen Li}, {and} \bibinfo{person}{Dilip Krishnan}.}
  \bibinfo{year}{2023}\natexlab{}.
\newblock \bibinfo{title}{Muse: Text-To-Image Generation via Masked Generative
  Transformers}.
\newblock
\newblock
\showeprint[arxiv]{2301.00704}~[cs.CV]


\bibitem[Chang et~al\mbox{.}(2002)]%
        {chang2002effect}
\bibfield{author}{\bibinfo{person}{Kuo-En Chang}, \bibinfo{person}{Yao-Ting
  Sung}, {and} \bibinfo{person}{Ine-Dai Chen}.}
  \bibinfo{year}{2002}\natexlab{}.
\newblock \showarticletitle{The effect of concept mapping to enhance text
  comprehension and summarization}.
\newblock \bibinfo{journal}{\emph{The Journal of Experimental Education}}
  \bibinfo{volume}{71}, \bibinfo{number}{1} (\bibinfo{year}{2002}),
  \bibinfo{pages}{5--23}.
\newblock


\bibitem[Chen(2010)]%
        {chen2010information}
\bibfield{author}{\bibinfo{person}{Chaomei Chen}.}
  \bibinfo{year}{2010}\natexlab{}.
\newblock \showarticletitle{Information visualization}.
\newblock \bibinfo{journal}{\emph{Wiley Interdisciplinary Reviews:
  Computational Statistics}} \bibinfo{volume}{2}, \bibinfo{number}{4}
  (\bibinfo{year}{2010}), \bibinfo{pages}{387--403}.
\newblock


\bibitem[Chen et~al\mbox{.}(2021)]%
        {chen2021evaluating}
\bibfield{author}{\bibinfo{person}{Mark Chen}, \bibinfo{person}{Jerry Tworek},
  \bibinfo{person}{Heewoo Jun}, \bibinfo{person}{Qiming Yuan},
  \bibinfo{person}{Henrique~Ponde de Oliveira~Pinto}, \bibinfo{person}{Jared
  Kaplan}, \bibinfo{person}{Harri Edwards}, \bibinfo{person}{Yuri Burda},
  \bibinfo{person}{Nicholas Joseph}, \bibinfo{person}{Greg Brockman},
  \bibinfo{person}{Alex Ray}, \bibinfo{person}{Raul Puri},
  \bibinfo{person}{Gretchen Krueger}, \bibinfo{person}{Michael Petrov},
  \bibinfo{person}{Heidy Khlaaf}, \bibinfo{person}{Girish Sastry},
  \bibinfo{person}{Pamela Mishkin}, \bibinfo{person}{Brooke Chan},
  \bibinfo{person}{Scott Gray}, \bibinfo{person}{Nick Ryder},
  \bibinfo{person}{Mikhail Pavlov}, \bibinfo{person}{Alethea Power},
  \bibinfo{person}{Lukasz Kaiser}, \bibinfo{person}{Mohammad Bavarian},
  \bibinfo{person}{Clemens Winter}, \bibinfo{person}{Philippe Tillet},
  \bibinfo{person}{Felipe~Petroski Such}, \bibinfo{person}{Dave Cummings},
  \bibinfo{person}{Matthias Plappert}, \bibinfo{person}{Fotios Chantzis},
  \bibinfo{person}{Elizabeth Barnes}, \bibinfo{person}{Ariel Herbert-Voss},
  \bibinfo{person}{William~Hebgen Guss}, \bibinfo{person}{Alex Nichol},
  \bibinfo{person}{Alex Paino}, \bibinfo{person}{Nikolas Tezak},
  \bibinfo{person}{Jie Tang}, \bibinfo{person}{Igor Babuschkin},
  \bibinfo{person}{Suchir Balaji}, \bibinfo{person}{Shantanu Jain},
  \bibinfo{person}{William Saunders}, \bibinfo{person}{Christopher Hesse},
  \bibinfo{person}{Andrew~N. Carr}, \bibinfo{person}{Jan Leike},
  \bibinfo{person}{Josh Achiam}, \bibinfo{person}{Vedant Misra},
  \bibinfo{person}{Evan Morikawa}, \bibinfo{person}{Alec Radford},
  \bibinfo{person}{Matthew Knight}, \bibinfo{person}{Miles Brundage},
  \bibinfo{person}{Mira Murati}, \bibinfo{person}{Katie Mayer},
  \bibinfo{person}{Peter Welinder}, \bibinfo{person}{Bob McGrew},
  \bibinfo{person}{Dario Amodei}, \bibinfo{person}{Sam McCandlish},
  \bibinfo{person}{Ilya Sutskever}, {and} \bibinfo{person}{Wojciech Zaremba}.}
  \bibinfo{year}{2021}\natexlab{}.
\newblock \bibinfo{title}{Evaluating Large Language Models Trained on Code}.
\newblock
\newblock
\showeprint[arxiv]{2107.03374}~[cs.LG]


\bibitem[Chen and Xia(2022)]%
        {crossdata2022}
\bibfield{author}{\bibinfo{person}{Zhutian Chen} {and} \bibinfo{person}{Haijun
  Xia}.} \bibinfo{year}{2022}\natexlab{}.
\newblock \showarticletitle{CrossData: Leveraging Text-Data Connections for
  Authoring Data Documents}. In \bibinfo{booktitle}{\emph{Proceedings of the
  2022 CHI Conference on Human Factors in Computing Systems}} (New Orleans, LA,
  USA) \emph{(\bibinfo{series}{CHI '22})}. \bibinfo{publisher}{Association for
  Computing Machinery}, \bibinfo{address}{New York, NY, USA}, Article
  \bibinfo{articleno}{95}, \bibinfo{numpages}{15}~pages.
\newblock
\showISBNx{9781450391573}
\urldef\tempurl%
\url{https://doi.org/10.1145/3491102.3517485}
\showDOI{\tempurl}


\bibitem[Cheng et~al\mbox{.}(2001)]%
        {cheng2001cognitive}
\bibfield{author}{\bibinfo{person}{Peter C.-H. Cheng}, \bibinfo{person}{Ric~K.
  Lowe}, {and} \bibinfo{person}{Mike Scaife}.} \bibinfo{year}{2001}\natexlab{}.
\newblock \bibinfo{booktitle}{\emph{Cognitive Science Approaches To
  Understanding Diagrammatic Representations}}.
\newblock \bibinfo{publisher}{Springer Netherlands},
  \bibinfo{address}{Dordrecht}, \bibinfo{pages}{79--94}.
\newblock
\showISBNx{978-94-017-3524-7}
\urldef\tempurl%
\url{https://doi.org/10.1007/978-94-017-3524-7_5}
\showDOI{\tempurl}


\bibitem[Chi et~al\mbox{.}(2021)]%
        {markdownvideo2021}
\bibfield{author}{\bibinfo{person}{Peggy Chi}, \bibinfo{person}{Nathan Frey},
  \bibinfo{person}{Katrina Panovich}, {and} \bibinfo{person}{Irfan Essa}.}
  \bibinfo{year}{2021}\natexlab{}.
\newblock \showarticletitle{Automatic Instructional Video Creation from a
  Markdown-Formatted Tutorial}. In \bibinfo{booktitle}{\emph{The 34th Annual
  ACM Symposium on User Interface Software and Technology}} (Virtual Event,
  USA) \emph{(\bibinfo{series}{UIST '21})}. \bibinfo{publisher}{Association for
  Computing Machinery}, \bibinfo{address}{New York, NY, USA},
  \bibinfo{pages}{677–690}.
\newblock
\showISBNx{9781450386357}
\urldef\tempurl%
\url{https://doi.org/10.1145/3472749.3474778}
\showDOI{\tempurl}


\bibitem[Chung et~al\mbox{.}(2022)]%
        {chung2022talebrush}
\bibfield{author}{\bibinfo{person}{John Joon~Young Chung},
  \bibinfo{person}{Wooseok Kim}, \bibinfo{person}{Kang~Min Yoo},
  \bibinfo{person}{Hwaran Lee}, \bibinfo{person}{Eytan Adar}, {and}
  \bibinfo{person}{Minsuk Chang}.} \bibinfo{year}{2022}\natexlab{}.
\newblock \showarticletitle{TaleBrush: Visual Sketching of Story Generation
  with Pretrained Language Models}. In \bibinfo{booktitle}{\emph{Extended
  Abstracts of the 2022 CHI Conference on Human Factors in Computing Systems}}
  (New Orleans, LA, USA) \emph{(\bibinfo{series}{CHI EA '22})}.
  \bibinfo{publisher}{Association for Computing Machinery},
  \bibinfo{address}{New York, NY, USA}, Article \bibinfo{articleno}{172},
  \bibinfo{numpages}{4}~pages.
\newblock
\showISBNx{9781450391566}
\urldef\tempurl%
\url{https://doi.org/10.1145/3491101.3519873}
\showDOI{\tempurl}


\bibitem[Clarinval et~al\mbox{.}(2018)]%
        {evoq2018}
\bibfield{author}{\bibinfo{person}{Antoine Clarinval},
  \bibinfo{person}{Isabelle Linden}, \bibinfo{person}{Anne Wallemacq}, {and}
  \bibinfo{person}{Bruno Dumas}.} \bibinfo{year}{2018}\natexlab{}.
\newblock \showarticletitle{Evoq: A Visualization Tool to Support Structural
  Analysis of Text Documents}. In \bibinfo{booktitle}{\emph{Proceedings of the
  ACM Symposium on Document Engineering 2018}} (Halifax, NS, Canada)
  \emph{(\bibinfo{series}{DocEng '18})}. \bibinfo{publisher}{Association for
  Computing Machinery}, \bibinfo{address}{New York, NY, USA}, Article
  \bibinfo{articleno}{27}, \bibinfo{numpages}{10}~pages.
\newblock
\showISBNx{9781450357692}
\urldef\tempurl%
\url{https://doi.org/10.1145/3209280.3209533}
\showDOI{\tempurl}


\bibitem[Clark and Paivio(1991)]%
        {clark1991dual}
\bibfield{author}{\bibinfo{person}{James~M Clark} {and} \bibinfo{person}{Allan
  Paivio}.} \bibinfo{year}{1991}\natexlab{}.
\newblock \showarticletitle{Dual coding theory and education}.
\newblock \bibinfo{journal}{\emph{Educational psychology review}}
  \bibinfo{volume}{3} (\bibinfo{year}{1991}), \bibinfo{pages}{149--210}.
\newblock


\bibitem[Cohen et~al\mbox{.}(1997)]%
        {cohen1997quickset}
\bibfield{author}{\bibinfo{person}{Philip~R. Cohen}, \bibinfo{person}{Michael
  Johnston}, \bibinfo{person}{David McGee}, \bibinfo{person}{Sharon Oviatt},
  \bibinfo{person}{Jay Pittman}, \bibinfo{person}{Ira Smith},
  \bibinfo{person}{Liang Chen}, {and} \bibinfo{person}{Josh Clow}.}
  \bibinfo{year}{1997}\natexlab{}.
\newblock \showarticletitle{QuickSet: Multimodal Interaction for Distributed
  Applications}. In \bibinfo{booktitle}{\emph{Proceedings of the Fifth ACM
  International Conference on Multimedia}} (Seattle, Washington, USA)
  \emph{(\bibinfo{series}{MULTIMEDIA '97})}. \bibinfo{publisher}{Association
  for Computing Machinery}, \bibinfo{address}{New York, NY, USA},
  \bibinfo{pages}{31–40}.
\newblock
\showISBNx{0897919912}
\urldef\tempurl%
\url{https://doi.org/10.1145/266180.266328}
\showDOI{\tempurl}


\bibitem[Cui et~al\mbox{.}(2019)]%
        {Text-to-viz2019}
\bibfield{author}{\bibinfo{person}{Weiwei Cui}, \bibinfo{person}{Xiaoyu Zhang},
  \bibinfo{person}{Yun Wang}, \bibinfo{person}{He Huang}, \bibinfo{person}{Bei
  Chen}, \bibinfo{person}{Lei Fang}, \bibinfo{person}{Haidong Zhang},
  \bibinfo{person}{Jian-Guan Lou}, {and} \bibinfo{person}{Dongmei Zhang}.}
  \bibinfo{year}{2019}\natexlab{}.
\newblock \showarticletitle{Text-to-viz: Automatic generation of infographics
  from proportion-related natural language statements}.
\newblock \bibinfo{journal}{\emph{IEEE transactions on visualization and
  computer graphics}} \bibinfo{volume}{26}, \bibinfo{number}{1}
  (\bibinfo{year}{2019}), \bibinfo{pages}{906--916}.
\newblock


\bibitem[Di~Fede et~al\mbox{.}(2022)]%
        {di22idea}
\bibfield{author}{\bibinfo{person}{Giulia Di~Fede}, \bibinfo{person}{Davide
  Rocchesso}, \bibinfo{person}{Steven~P. Dow}, {and} \bibinfo{person}{Salvatore
  Andolina}.} \bibinfo{year}{2022}\natexlab{}.
\newblock \showarticletitle{The Idea Machine: LLM-Based Expansion, Rewriting,
  Combination, and Suggestion of Ideas}. In
  \bibinfo{booktitle}{\emph{Proceedings of the 14th Conference on Creativity
  and Cognition}} (Venice, Italy) \emph{(\bibinfo{series}{C\&C '22})}.
  \bibinfo{publisher}{Association for Computing Machinery},
  \bibinfo{address}{New York, NY, USA}, \bibinfo{pages}{623–627}.
\newblock
\showISBNx{9781450393270}
\urldef\tempurl%
\url{https://doi.org/10.1145/3527927.3535197}
\showDOI{\tempurl}


\bibitem[Esser et~al\mbox{.}(2023)]%
        {esser2023structure}
\bibfield{author}{\bibinfo{person}{Patrick Esser}, \bibinfo{person}{Johnathan
  Chiu}, \bibinfo{person}{Parmida Atighehchian}, \bibinfo{person}{Jonathan
  Granskog}, {and} \bibinfo{person}{Anastasis Germanidis}.}
  \bibinfo{year}{2023}\natexlab{}.
\newblock \bibinfo{title}{Structure and Content-Guided Video Synthesis with
  Diffusion Models}.
\newblock
\newblock
\showeprint[arxiv]{2302.03011}~[cs.CV]


\bibitem[Fast et~al\mbox{.}(2018)]%
        {iris2018}
\bibfield{author}{\bibinfo{person}{Ethan Fast}, \bibinfo{person}{Binbin Chen},
  \bibinfo{person}{Julia Mendelsohn}, \bibinfo{person}{Jonathan Bassen}, {and}
  \bibinfo{person}{Michael~S. Bernstein}.} \bibinfo{year}{2018}\natexlab{}.
\newblock \showarticletitle{Iris: A Conversational Agent for Complex Tasks}. In
  \bibinfo{booktitle}{\emph{Proceedings of the 2018 CHI Conference on Human
  Factors in Computing Systems}} (Montreal QC, Canada)
  \emph{(\bibinfo{series}{CHI '18})}. \bibinfo{publisher}{Association for
  Computing Machinery}, \bibinfo{address}{New York, NY, USA},
  \bibinfo{pages}{1–12}.
\newblock
\showISBNx{9781450356206}
\urldef\tempurl%
\url{https://doi.org/10.1145/3173574.3174047}
\showDOI{\tempurl}


\bibitem[Fatemeh et~al\mbox{.}(2011)]%
        {fatemeh2011icmap}
\bibfield{author}{\bibinfo{person}{Hendijanifard Fatemeh},
  \bibinfo{person}{Kardan Ahmad}, {and} \bibinfo{person}{Dibay~Moghadam
  Mohammad}.} \bibinfo{year}{2011}\natexlab{}.
\newblock \showarticletitle{ICMAP: An interactive tool for concept map
  generation to facilitate learning process}.
\newblock \bibinfo{journal}{\emph{Procedia Computer Science}}
  \bibinfo{volume}{3} (\bibinfo{year}{2011}), \bibinfo{pages}{524--529}.
\newblock


\bibitem[Gao et~al\mbox{.}(2015)]%
        {datatone}
\bibfield{author}{\bibinfo{person}{Tong Gao}, \bibinfo{person}{Mira Dontcheva},
  \bibinfo{person}{Eytan Adar}, \bibinfo{person}{Zhicheng Liu}, {and}
  \bibinfo{person}{Karrie~G. Karahalios}.} \bibinfo{year}{2015}\natexlab{}.
\newblock \showarticletitle{DataTone: Managing Ambiguity in Natural Language
  Interfaces for Data Visualization}. In \bibinfo{booktitle}{\emph{Proceedings
  of the 28th Annual ACM Symposium on User Interface Software \& Technology}}
  (Charlotte, NC, USA) \emph{(\bibinfo{series}{UIST '15})}.
  \bibinfo{publisher}{Association for Computing Machinery},
  \bibinfo{address}{New York, NY, USA}, \bibinfo{pages}{489–500}.
\newblock
\showISBNx{9781450337793}
\urldef\tempurl%
\url{https://doi.org/10.1145/2807442.2807478}
\showDOI{\tempurl}


\bibitem[Gero et~al\mbox{.}(2022)]%
        {gero2022sparks}
\bibfield{author}{\bibinfo{person}{Katy~Ilonka Gero}, \bibinfo{person}{Vivian
  Liu}, {and} \bibinfo{person}{Lydia Chilton}.}
  \bibinfo{year}{2022}\natexlab{}.
\newblock \showarticletitle{Sparks: Inspiration for Science Writing Using
  Language Models}. In \bibinfo{booktitle}{\emph{Designing Interactive Systems
  Conference}} (Virtual Event, Australia) \emph{(\bibinfo{series}{DIS '22})}.
  \bibinfo{publisher}{Association for Computing Machinery},
  \bibinfo{address}{New York, NY, USA}, \bibinfo{pages}{1002–1019}.
\newblock
\showISBNx{9781450393584}
\urldef\tempurl%
\url{https://doi.org/10.1145/3532106.3533533}
\showDOI{\tempurl}


\bibitem[Goodwin(2015)]%
        {goodwin2015professional}
\bibfield{author}{\bibinfo{person}{Charles Goodwin}.}
  \bibinfo{year}{2015}\natexlab{}.
\newblock \bibinfo{booktitle}{\emph{Professional Vision}}.
\newblock \bibinfo{publisher}{Springer Fachmedien Wiesbaden},
  \bibinfo{address}{Wiesbaden}, \bibinfo{pages}{387--425}.
\newblock
\showISBNx{978-3-531-19381-6}
\urldef\tempurl%
\url{https://doi.org/10.1007/978-3-531-19381-6_20}
\showDOI{\tempurl}


\bibitem[Guastello et~al\mbox{.}(1989)]%
        {guastello1989verbal}
\bibfield{author}{\bibinfo{person}{Stephen~J Guastello}, \bibinfo{person}{Mary
  Traut}, {and} \bibinfo{person}{Gene Korienek}.}
  \bibinfo{year}{1989}\natexlab{}.
\newblock \showarticletitle{Verbal versus pictorial representations of objects
  in a human-computer interface}.
\newblock \bibinfo{journal}{\emph{International journal of man-machine
  studies}} \bibinfo{volume}{31}, \bibinfo{number}{1} (\bibinfo{year}{1989}),
  \bibinfo{pages}{99--120}.
\newblock


\bibitem[Hahn and Kim(1999)]%
        {kim1999why}
\bibfield{author}{\bibinfo{person}{Jungpil Hahn} {and} \bibinfo{person}{Jinwoo
  Kim}.} \bibinfo{year}{1999}\natexlab{}.
\newblock \showarticletitle{Why Are Some Diagrams Easier to Work with? Effects
  of Diagrammatic Representation on the Cognitive Intergration Process of
  Systems Analysis and Design}.
\newblock \bibinfo{journal}{\emph{ACM Trans. Comput.-Hum. Interact.}}
  \bibinfo{volume}{6}, \bibinfo{number}{3} (\bibinfo{date}{sep}
  \bibinfo{year}{1999}), \bibinfo{pages}{181–213}.
\newblock
\showISSN{1073-0516}
\urldef\tempurl%
\url{https://doi.org/10.1145/329693.329694}
\showDOI{\tempurl}


\bibitem[Hayama and Sato(2020)]%
        {videoconceptmapgen2020}
\bibfield{author}{\bibinfo{person}{Tessai Hayama} {and} \bibinfo{person}{Shuma
  Sato}.} \bibinfo{year}{2020}\natexlab{}.
\newblock \showarticletitle{Supporting Online Video e-Learning with
  Semi-automatic Concept-Map Generation}. In \bibinfo{booktitle}{\emph{Learning
  and Collaboration Technologies. Designing, Developing and Deploying Learning
  Experiences}}. \bibinfo{publisher}{Springer International Publishing},
  \bibinfo{address}{Cham}, \bibinfo{pages}{64--76}.
\newblock
\showISBNx{978-3-030-50513-4}


\bibitem[Hayatpur et~al\mbox{.}(2023)]%
        {crosscode}
\bibfield{author}{\bibinfo{person}{Devamardeep Hayatpur},
  \bibinfo{person}{Daniel Wigdor}, {and} \bibinfo{person}{Haijun Xia}.}
  \bibinfo{year}{2023}\natexlab{}.
\newblock \showarticletitle{CrossCode: Multi-Level Visualization of Program
  Execution}. In \bibinfo{booktitle}{\emph{Proceedings of the 2023 CHI
  Conference on Human Factors in Computing Systems}} (Hamburg, Germany)
  \emph{(\bibinfo{series}{CHI '23})}. \bibinfo{publisher}{Association for
  Computing Machinery}, \bibinfo{address}{New York, NY, USA}, Article
  \bibinfo{articleno}{593}, \bibinfo{numpages}{13}~pages.
\newblock
\showISBNx{9781450394215}
\urldef\tempurl%
\url{https://doi.org/10.1145/3544548.3581390}
\showDOI{\tempurl}


\bibitem[Hearst and Tory(2019)]%
        {r40}
\bibfield{author}{\bibinfo{person}{Marti Hearst} {and} \bibinfo{person}{Melanie
  Tory}.} \bibinfo{year}{2019}\natexlab{}.
\newblock \showarticletitle{Would You Like A Chart With That? Incorporating
  Visualizations into Conversational Interfaces}. In
  \bibinfo{booktitle}{\emph{2019 IEEE Visualization Conference (VIS)}}.
  \bibinfo{publisher}{IEEE}, \bibinfo{pages}{1--5}.
\newblock
\urldef\tempurl%
\url{https://doi.org/10.1109/VISUAL.2019.8933766}
\showDOI{\tempurl}


\bibitem[Hong et~al\mbox{.}(2022)]%
        {hong2022avatarclip}
\bibfield{author}{\bibinfo{person}{Fangzhou Hong}, \bibinfo{person}{Mingyuan
  Zhang}, \bibinfo{person}{Liang Pan}, \bibinfo{person}{Zhongang Cai},
  \bibinfo{person}{Lei Yang}, {and} \bibinfo{person}{Ziwei Liu}.}
  \bibinfo{year}{2022}\natexlab{}.
\newblock \bibinfo{title}{AvatarCLIP: Zero-Shot Text-Driven Generation and
  Animation of 3D Avatars}.
\newblock
\newblock
\showeprint[arxiv]{2205.08535}~[cs.CV]


\bibitem[Hu et~al\mbox{.}(2017)]%
        {socialmediatextviz2017}
\bibfield{author}{\bibinfo{person}{Mengdie Hu}, \bibinfo{person}{Krist
  Wongsuphasawat}, {and} \bibinfo{person}{John Stasko}.}
  \bibinfo{year}{2017}\natexlab{}.
\newblock \showarticletitle{Visualizing Social Media Content with SentenTree}.
\newblock \bibinfo{journal}{\emph{IEEE Transactions on Visualization and
  Computer Graphics}} \bibinfo{volume}{23}, \bibinfo{number}{1}
  (\bibinfo{year}{2017}), \bibinfo{pages}{621--630}.
\newblock
\urldef\tempurl%
\url{https://doi.org/10.1109/TVCG.2016.2598590}
\showDOI{\tempurl}


\bibitem[Hwang et~al\mbox{.}(2011)]%
        {conceptmapmobile2011}
\bibfield{author}{\bibinfo{person}{Gwo-Jen Hwang}, \bibinfo{person}{Po-Han Wu},
  {and} \bibinfo{person}{Hui-Ru Ke}.} \bibinfo{year}{2011}\natexlab{}.
\newblock \showarticletitle{An interactive concept map approach to supporting
  mobile learning activities for natural science courses}.
\newblock \bibinfo{journal}{\emph{Computers \& education}}
  \bibinfo{volume}{57}, \bibinfo{number}{4} (\bibinfo{year}{2011}),
  \bibinfo{pages}{2272--2280}.
\newblock


\bibitem[Hwang et~al\mbox{.}(2013)]%
        {hwang2013concept}
\bibfield{author}{\bibinfo{person}{Gwo-Jen Hwang}, \bibinfo{person}{Li-Hsueh
  Yang}, {and} \bibinfo{person}{Sheng-Yuan Wang}.}
  \bibinfo{year}{2013}\natexlab{}.
\newblock \showarticletitle{A concept map-embedded educational computer game
  for improving students' learning performance in natural science courses}.
\newblock \bibinfo{journal}{\emph{Computers \& Education}}
  \bibinfo{volume}{69} (\bibinfo{year}{2013}), \bibinfo{pages}{121--130}.
\newblock


\bibitem[Jian and Wu(2015)]%
        {jian2015using}
\bibfield{author}{\bibinfo{person}{Yu-Cin Jian} {and}
  \bibinfo{person}{Chao-Jung Wu}.} \bibinfo{year}{2015}\natexlab{}.
\newblock \showarticletitle{Using eye tracking to investigate semantic and
  spatial representations of scientific diagrams during text-diagram
  integration}.
\newblock \bibinfo{journal}{\emph{Journal of Science Education and Technology}}
   \bibinfo{volume}{24} (\bibinfo{year}{2015}), \bibinfo{pages}{43--55}.
\newblock


\bibitem[Jiang et~al\mbox{.}(2023)]%
        {jiang2023log}
\bibfield{author}{\bibinfo{person}{Peiling Jiang}, \bibinfo{person}{Fuling
  Sun}, {and} \bibinfo{person}{Haijun Xia}.} \bibinfo{year}{2023}\natexlab{}.
\newblock \showarticletitle{Log-It: Supporting Programming with Interactive,
  Contextual, Structured, and Visual Logs}. In
  \bibinfo{booktitle}{\emph{Proceedings of the 2023 CHI Conference on Human
  Factors in Computing Systems}} (Hamburg, Germany) \emph{(\bibinfo{series}{CHI
  '23})}. \bibinfo{publisher}{Association for Computing Machinery},
  \bibinfo{address}{New York, NY, USA}, Article \bibinfo{articleno}{594},
  \bibinfo{numpages}{16}~pages.
\newblock
\showISBNx{9781450394215}
\urldef\tempurl%
\url{https://doi.org/10.1145/3544548.3581403}
\showDOI{\tempurl}


\bibitem[Kim et~al\mbox{.}(2018)]%
        {linkingtxt&tables2018}
\bibfield{author}{\bibinfo{person}{Dae~Hyun Kim}, \bibinfo{person}{Enamul
  Hoque}, \bibinfo{person}{Juho Kim}, {and} \bibinfo{person}{Maneesh
  Agrawala}.} \bibinfo{year}{2018}\natexlab{}.
\newblock \showarticletitle{Facilitating Document Reading by Linking Text and
  Tables}. In \bibinfo{booktitle}{\emph{Proceedings of the 31st Annual ACM
  Symposium on User Interface Software and Technology}} (Berlin, Germany)
  \emph{(\bibinfo{series}{UIST '18})}. \bibinfo{publisher}{Association for
  Computing Machinery}, \bibinfo{address}{New York, NY, USA},
  \bibinfo{pages}{423–434}.
\newblock
\showISBNx{9781450359481}
\urldef\tempurl%
\url{https://doi.org/10.1145/3242587.3242617}
\showDOI{\tempurl}


\bibitem[Kim et~al\mbox{.}(2019)]%
        {voicecuts}
\bibfield{author}{\bibinfo{person}{Yea-Seul Kim}, \bibinfo{person}{Mira
  Dontcheva}, \bibinfo{person}{Eytan Adar}, {and} \bibinfo{person}{Jessica
  Hullman}.} \bibinfo{year}{2019}\natexlab{}.
\newblock \showarticletitle{Vocal Shortcuts for Creative Experts}. In
  \bibinfo{booktitle}{\emph{Proceedings of the 2019 CHI Conference on Human
  Factors in Computing Systems}} (Glasgow, Scotland Uk)
  \emph{(\bibinfo{series}{CHI '19})}. \bibinfo{publisher}{Association for
  Computing Machinery}, \bibinfo{address}{New York, NY, USA},
  \bibinfo{pages}{1–14}.
\newblock
\showISBNx{9781450359702}
\urldef\tempurl%
\url{https://doi.org/10.1145/3290605.3300562}
\showDOI{\tempurl}


\bibitem[Kung et~al\mbox{.}(2023)]%
        {kung2023performance}
\bibfield{author}{\bibinfo{person}{Tiffany~H Kung}, \bibinfo{person}{Morgan
  Cheatham}, \bibinfo{person}{Arielle Medenilla}, \bibinfo{person}{Czarina
  Sillos}, \bibinfo{person}{Lorie De~Leon}, \bibinfo{person}{Camille
  Elepa{\~n}o}, \bibinfo{person}{Maria Madriaga}, \bibinfo{person}{Rimel
  Aggabao}, \bibinfo{person}{Giezel Diaz-Candido}, \bibinfo{person}{James
  Maningo}, {et~al\mbox{.}}} \bibinfo{year}{2023}\natexlab{}.
\newblock \showarticletitle{Performance of ChatGPT on USMLE: Potential for
  AI-assisted medical education using large language models}.
\newblock \bibinfo{journal}{\emph{PLoS digital health}} \bibinfo{volume}{2},
  \bibinfo{number}{2} (\bibinfo{year}{2023}), \bibinfo{pages}{e0000198}.
\newblock


\bibitem[Laban et~al\mbox{.}(2021)]%
        {laban2021simple}
\bibfield{author}{\bibinfo{person}{Philippe Laban}, \bibinfo{person}{Tobias
  Schnabel}, \bibinfo{person}{Paul Bennett}, {and} \bibinfo{person}{Marti~A.
  Hearst}.} \bibinfo{year}{2021}\natexlab{}.
\newblock \bibinfo{title}{Keep it Simple: Unsupervised Simplification of
  Multi-Paragraph Text}.
\newblock
\newblock
\showeprint[arxiv]{2107.03444}~[cs.CL]


\bibitem[Larkin and Simon(1987)]%
        {larkin1987diagram}
\bibfield{author}{\bibinfo{person}{Jill~H Larkin} {and}
  \bibinfo{person}{Herbert~A Simon}.} \bibinfo{year}{1987}\natexlab{}.
\newblock \showarticletitle{Why a diagram is (sometimes) worth ten thousand
  words}.
\newblock \bibinfo{journal}{\emph{Cognitive science}} \bibinfo{volume}{11},
  \bibinfo{number}{1} (\bibinfo{year}{1987}), \bibinfo{pages}{65--100}.
\newblock


\bibitem[Liu et~al\mbox{.}(2018)]%
        {conceptscape2018}
\bibfield{author}{\bibinfo{person}{Ching Liu}, \bibinfo{person}{Juho Kim},
  {and} \bibinfo{person}{Hao-Chuan Wang}.} \bibinfo{year}{2018}\natexlab{}.
\newblock \showarticletitle{ConceptScape: Collaborative Concept Mapping for
  Video Learning}. In \bibinfo{booktitle}{\emph{Proceedings of the 2018 CHI
  Conference on Human Factors in Computing Systems}} (Montreal QC, Canada)
  \emph{(\bibinfo{series}{CHI '18})}. \bibinfo{publisher}{Association for
  Computing Machinery}, \bibinfo{address}{New York, NY, USA},
  \bibinfo{pages}{1–12}.
\newblock
\showISBNx{9781450356206}
\urldef\tempurl%
\url{https://doi.org/10.1145/3173574.3173961}
\showDOI{\tempurl}


\bibitem[Liu et~al\mbox{.}(2022)]%
        {liu2022wig}
\bibfield{author}{\bibinfo{person}{Michael~Xieyang Liu},
  \bibinfo{person}{Andrew Kuznetsov}, \bibinfo{person}{Yongsung Kim},
  \bibinfo{person}{Joseph~Chee Chang}, \bibinfo{person}{Aniket Kittur}, {and}
  \bibinfo{person}{Brad~A. Myers}.} \bibinfo{year}{2022}\natexlab{}.
\newblock \showarticletitle{Wigglite: Low-Cost Information Collection and
  Triage}. In \bibinfo{booktitle}{\emph{Proceedings of the 35th Annual ACM
  Symposium on User Interface Software and Technology}} (Bend, OR, USA)
  \emph{(\bibinfo{series}{UIST '22})}. \bibinfo{publisher}{Association for
  Computing Machinery}, \bibinfo{address}{New York, NY, USA}, Article
  \bibinfo{articleno}{32}, \bibinfo{numpages}{16}~pages.
\newblock
\showISBNx{9781450393201}
\urldef\tempurl%
\url{https://doi.org/10.1145/3526113.3545661}
\showDOI{\tempurl}


\bibitem[Liu and Chilton(2022)]%
        {designguideprompt2022}
\bibfield{author}{\bibinfo{person}{Vivian Liu} {and} \bibinfo{person}{Lydia~B
  Chilton}.} \bibinfo{year}{2022}\natexlab{}.
\newblock \showarticletitle{Design Guidelines for Prompt Engineering
  Text-to-Image Generative Models}. In \bibinfo{booktitle}{\emph{Proceedings of
  the 2022 CHI Conference on Human Factors in Computing Systems}} (New Orleans,
  LA, USA) \emph{(\bibinfo{series}{CHI '22})}. \bibinfo{publisher}{Association
  for Computing Machinery}, \bibinfo{address}{New York, NY, USA}, Article
  \bibinfo{articleno}{384}, \bibinfo{numpages}{23}~pages.
\newblock
\showISBNx{9781450391573}
\urldef\tempurl%
\url{https://doi.org/10.1145/3491102.3501825}
\showDOI{\tempurl}


\bibitem[MacNeil et~al\mbox{.}(2023)]%
        {macneil2022automatically}
\bibfield{author}{\bibinfo{person}{Stephen MacNeil}, \bibinfo{person}{Andrew
  Tran}, \bibinfo{person}{Juho Leinonen}, \bibinfo{person}{Paul Denny},
  \bibinfo{person}{Joanne Kim}, \bibinfo{person}{Arto Hellas},
  \bibinfo{person}{Seth Bernstein}, {and} \bibinfo{person}{Sami Sarsa}.}
  \bibinfo{year}{2023}\natexlab{}.
\newblock \showarticletitle{Automatically Generating CS Learning Materials with
  Large Language Models}. In \bibinfo{booktitle}{\emph{Proceedings of the 54th
  ACM Technical Symposium on Computer Science Education V. 2}} (Toronto ON,
  Canada) \emph{(\bibinfo{series}{SIGCSE 2023})}.
  \bibinfo{publisher}{Association for Computing Machinery},
  \bibinfo{address}{New York, NY, USA}, \bibinfo{pages}{1176}.
\newblock
\showISBNx{9781450394338}
\urldef\tempurl%
\url{https://doi.org/10.1145/3545947.3569630}
\showDOI{\tempurl}


\bibitem[McClellan et~al\mbox{.}(2004)]%
        {CNT2004}
\bibfield{author}{\bibinfo{person}{J.H. McClellan}, \bibinfo{person}{L.D.
  Harvel}, \bibinfo{person}{R. Velmurugan}, \bibinfo{person}{M. Borkar}, {and}
  \bibinfo{person}{C. Scheibe}.} \bibinfo{year}{2004}\natexlab{}.
\newblock \showarticletitle{CNT: concept-map based navigation and discovery in
  a repository of learning content}. In \bibinfo{booktitle}{\emph{34th Annual
  Frontiers in Education, 2004. FIE 2004.}} \bibinfo{publisher}{IEEE},
  \bibinfo{pages}{F1F--13}.
\newblock
\urldef\tempurl%
\url{https://doi.org/10.1109/FIE.2004.1408581}
\showDOI{\tempurl}


\bibitem[Metoyer et~al\mbox{.}(2018)]%
        {couplingstory2viz2018}
\bibfield{author}{\bibinfo{person}{Ronald Metoyer}, \bibinfo{person}{Qiyu Zhi},
  \bibinfo{person}{Bart Janczuk}, {and} \bibinfo{person}{Walter Scheirer}.}
  \bibinfo{year}{2018}\natexlab{}.
\newblock \showarticletitle{Coupling Story to Visualization: Using Textual
  Analysis as a Bridge Between Data and Interpretation}. In
  \bibinfo{booktitle}{\emph{23rd International Conference on Intelligent User
  Interfaces}} (Tokyo, Japan) \emph{(\bibinfo{series}{IUI '18})}.
  \bibinfo{publisher}{Association for Computing Machinery},
  \bibinfo{address}{New York, NY, USA}, \bibinfo{pages}{503–507}.
\newblock
\showISBNx{9781450349451}
\urldef\tempurl%
\url{https://doi.org/10.1145/3172944.3173007}
\showDOI{\tempurl}


\bibitem[More and Phalnikar(2012)]%
        {more2012generating}
\bibfield{author}{\bibinfo{person}{Priyanka More} {and} \bibinfo{person}{Rashmi
  Phalnikar}.} \bibinfo{year}{2012}\natexlab{}.
\newblock \showarticletitle{Generating UML diagrams from natural language
  specifications}.
\newblock \bibinfo{journal}{\emph{International Journal of Applied Information
  Systems}} \bibinfo{volume}{1}, \bibinfo{number}{8} (\bibinfo{year}{2012}),
  \bibinfo{pages}{19--23}.
\newblock


\bibitem[Moura et~al\mbox{.}(2017)]%
        {moura2017learning}
\bibfield{author}{\bibinfo{person}{Raphael Moura}, \bibinfo{person}{Michael
  Beer}, \bibinfo{person}{Edoardo Patelli}, {and} \bibinfo{person}{John
  Lewis}.} \bibinfo{year}{2017}\natexlab{}.
\newblock \showarticletitle{Learning from major accidents: Graphical
  representation and analysis of multi-attribute events to enhance risk
  communication}.
\newblock \bibinfo{journal}{\emph{Safety science}}  \bibinfo{volume}{99}
  (\bibinfo{year}{2017}), \bibinfo{pages}{58--70}.
\newblock


\bibitem[Màrquez et~al\mbox{.}(2008)]%
        {marquez2008semantic}
\bibfield{author}{\bibinfo{person}{Lluís Màrquez}, \bibinfo{person}{Xavier
  Carreras}, \bibinfo{person}{Kenneth~C. Litkowski}, {and}
  \bibinfo{person}{Suzanne Stevenson}.} \bibinfo{year}{2008}\natexlab{}.
\newblock \showarticletitle{{Semantic Role Labeling: An Introduction to the
  Special Issue}}.
\newblock \bibinfo{journal}{\emph{Computational Linguistics}}
  \bibinfo{volume}{34}, \bibinfo{number}{2} (\bibinfo{date}{06}
  \bibinfo{year}{2008}), \bibinfo{pages}{145--159}.
\newblock
\showISSN{0891-2017}
\urldef\tempurl%
\url{https://doi.org/10.1162/coli.2008.34.2.145}
\showDOI{\tempurl}
\showeprint{https://direct.mit.edu/coli/article-pdf/34/2/145/1798596/coli.2008.34.2.145.pdf}


\bibitem[Narechania et~al\mbox{.}(2020)]%
        {NL4DV2020}
\bibfield{author}{\bibinfo{person}{Arpit Narechania}, \bibinfo{person}{Arjun
  Srinivasan}, {and} \bibinfo{person}{John Stasko}.}
  \bibinfo{year}{2020}\natexlab{}.
\newblock \showarticletitle{NL4DV: A toolkit for generating analytic
  specifications for data visualization from natural language queries}.
\newblock \bibinfo{journal}{\emph{IEEE Transactions on Visualization and
  Computer Graphics}} \bibinfo{volume}{27}, \bibinfo{number}{2}
  (\bibinfo{year}{2020}), \bibinfo{pages}{369--379}.
\newblock


\bibitem[Norman and Draper(1986)]%
        {normancentered1986}
\bibfield{author}{\bibinfo{person}{Donald~A. Norman} {and}
  \bibinfo{person}{Stephen~W. Draper}.} \bibinfo{year}{1986}\natexlab{}.
\newblock \bibinfo{booktitle}{\emph{User Centered System Design; New
  Perspectives on Human-Computer Interaction}}.
\newblock \bibinfo{publisher}{L. Erlbaum Associates Inc.},
  \bibinfo{address}{USA}.
\newblock
\showISBNx{0898597811}


\bibitem[OpenAI(2023)]%
        {gpt4technical}
\bibfield{author}{\bibinfo{person}{OpenAI}.} \bibinfo{year}{2023}\natexlab{}.
\newblock \bibinfo{title}{GPT-4 Technical Report}.
\newblock
\newblock
\showeprint[arxiv]{2303.08774}~[cs.CL]


\bibitem[Palani et~al\mbox{.}(2022)]%
        {palani2022interweave}
\bibfield{author}{\bibinfo{person}{Srishti Palani}, \bibinfo{person}{Yingyi
  Zhou}, \bibinfo{person}{Sheldon Zhu}, {and} \bibinfo{person}{Steven~P. Dow}.}
  \bibinfo{year}{2022}\natexlab{}.
\newblock \showarticletitle{InterWeave: Presenting Search Suggestions in
  Context Scaffolds Information Search and Synthesis}. In
  \bibinfo{booktitle}{\emph{Proceedings of the 35th Annual ACM Symposium on
  User Interface Software and Technology}} (Bend, OR, USA)
  \emph{(\bibinfo{series}{UIST '22})}. \bibinfo{publisher}{Association for
  Computing Machinery}, \bibinfo{address}{New York, NY, USA}, Article
  \bibinfo{articleno}{93}, \bibinfo{numpages}{16}~pages.
\newblock
\showISBNx{9781450393201}
\urldef\tempurl%
\url{https://doi.org/10.1145/3526113.3545696}
\showDOI{\tempurl}


\bibitem[Palmer et~al\mbox{.}(2011)]%
        {palmer2011semantic}
\bibfield{author}{\bibinfo{person}{M. Palmer}, \bibinfo{person}{D. Gildea},
  {and} \bibinfo{person}{N. Xue}.} \bibinfo{year}{2011}\natexlab{}.
\newblock \bibinfo{booktitle}{\emph{Semantic Role Labeling}}.
\newblock \bibinfo{publisher}{Morgan \& Claypool Publishers}.
\newblock
\showISBNx{9781598298321}
\urldef\tempurl%
\url{https://books.google.com/books?id=saBdAQAAQBAJ}
\showURL{%
\tempurl}


\bibitem[Pearce et~al\mbox{.}(2022)]%
        {pearce2022examining}
\bibfield{author}{\bibinfo{person}{Hammond Pearce}, \bibinfo{person}{Benjamin
  Tan}, \bibinfo{person}{Baleegh Ahmad}, \bibinfo{person}{Ramesh Karri}, {and}
  \bibinfo{person}{Brendan Dolan-Gavitt}.} \bibinfo{year}{2022}\natexlab{}.
\newblock \bibinfo{title}{Examining Zero-Shot Vulnerability Repair with Large
  Language Models}.
\newblock
\newblock
\showeprint[arxiv]{2112.02125}~[cs.CR]


\bibitem[Piskorski and Yangarber(2013)]%
        {piskorski2013information}
\bibfield{author}{\bibinfo{person}{Jakub Piskorski} {and}
  \bibinfo{person}{Roman Yangarber}.} \bibinfo{year}{2013}\natexlab{}.
\newblock \bibinfo{booktitle}{\emph{Information Extraction: Past, Present and
  Future}}.
\newblock \bibinfo{publisher}{Springer Berlin Heidelberg},
  \bibinfo{address}{Berlin, Heidelberg}, \bibinfo{pages}{23--49}.
\newblock
\showISBNx{978-3-642-28569-1}
\urldef\tempurl%
\url{https://doi.org/10.1007/978-3-642-28569-1_2}
\showDOI{\tempurl}


\bibitem[Qin et~al\mbox{.}(2023)]%
        {qin2023chatgpt}
\bibfield{author}{\bibinfo{person}{Chengwei Qin}, \bibinfo{person}{Aston
  Zhang}, \bibinfo{person}{Zhuosheng Zhang}, \bibinfo{person}{Jiaao Chen},
  \bibinfo{person}{Michihiro Yasunaga}, {and} \bibinfo{person}{Diyi Yang}.}
  \bibinfo{year}{2023}\natexlab{}.
\newblock \bibinfo{title}{Is ChatGPT a General-Purpose Natural Language
  Processing Task Solver?}
\newblock
\newblock
\showeprint[arxiv]{2302.06476}~[cs.CL]


\bibitem[Reynolds and McDonell(2021)]%
        {reynolds2021prompt}
\bibfield{author}{\bibinfo{person}{Laria Reynolds} {and} \bibinfo{person}{Kyle
  McDonell}.} \bibinfo{year}{2021}\natexlab{}.
\newblock \bibinfo{title}{Prompt Programming for Large Language Models: Beyond
  the Few-Shot Paradigm}.
\newblock
\newblock
\showeprint[arxiv]{2102.07350}~[cs.CL]


\bibitem[Schwab et~al\mbox{.}(2017)]%
        {booc.io}
\bibfield{author}{\bibinfo{person}{Michail Schwab}, \bibinfo{person}{Hendrik
  Strobelt}, \bibinfo{person}{James Tompkin}, \bibinfo{person}{Colin
  Fredericks}, \bibinfo{person}{Connor Huff}, \bibinfo{person}{Dana Higgins},
  \bibinfo{person}{Anton Strezhnev}, \bibinfo{person}{Mayya Komisarchik},
  \bibinfo{person}{Gary King}, {and} \bibinfo{person}{Hanspeter Pfister}.}
  \bibinfo{year}{2017}\natexlab{}.
\newblock \showarticletitle{booc.io: An Education System with Hierarchical
  Concept Maps and Dynamic Non-linear Learning Plans}.
\newblock \bibinfo{journal}{\emph{IEEE Transactions on Visualization and
  Computer Graphics}} \bibinfo{volume}{23}, \bibinfo{number}{1}
  (\bibinfo{year}{2017}), \bibinfo{pages}{571--580}.
\newblock
\urldef\tempurl%
\url{https://doi.org/10.1109/TVCG.2016.2598518}
\showDOI{\tempurl}


\bibitem[Shinn et~al\mbox{.}(2023)]%
        {shinn2023reflexion}
\bibfield{author}{\bibinfo{person}{Noah Shinn}, \bibinfo{person}{Beck Labash},
  {and} \bibinfo{person}{Ashwin Gopinath}.} \bibinfo{year}{2023}\natexlab{}.
\newblock \bibinfo{title}{Reflexion: an autonomous agent with dynamic memory
  and self-reflection}.
\newblock
\newblock
\showeprint[arxiv]{2303.11366}~[cs.AI]


\bibitem[Singer et~al\mbox{.}(2022)]%
        {singer2022makeavideo}
\bibfield{author}{\bibinfo{person}{Uriel Singer}, \bibinfo{person}{Adam
  Polyak}, \bibinfo{person}{Thomas Hayes}, \bibinfo{person}{Xi Yin},
  \bibinfo{person}{Jie An}, \bibinfo{person}{Songyang Zhang},
  \bibinfo{person}{Qiyuan Hu}, \bibinfo{person}{Harry Yang},
  \bibinfo{person}{Oron Ashual}, \bibinfo{person}{Oran Gafni},
  \bibinfo{person}{Devi Parikh}, \bibinfo{person}{Sonal Gupta}, {and}
  \bibinfo{person}{Yaniv Taigman}.} \bibinfo{year}{2022}\natexlab{}.
\newblock \bibinfo{title}{Make-A-Video: Text-to-Video Generation without
  Text-Video Data}.
\newblock
\newblock
\showeprint[arxiv]{2209.14792}~[cs.CV]


\bibitem[Singh et~al\mbox{.}(2022)]%
        {singh2022progprompt}
\bibfield{author}{\bibinfo{person}{Ishika Singh}, \bibinfo{person}{Valts
  Blukis}, \bibinfo{person}{Arsalan Mousavian}, \bibinfo{person}{Ankit Goyal},
  \bibinfo{person}{Danfei Xu}, \bibinfo{person}{Jonathan Tremblay},
  \bibinfo{person}{Dieter Fox}, \bibinfo{person}{Jesse Thomason}, {and}
  \bibinfo{person}{Animesh Garg}.} \bibinfo{year}{2022}\natexlab{}.
\newblock \bibinfo{title}{ProgPrompt: Generating Situated Robot Task Plans
  using Large Language Models}.
\newblock
\newblock
\showeprint[arxiv]{2209.11302}~[cs.RO]


\bibitem[Spoerri(1993)]%
        {spoerri1993infocrystal}
\bibfield{author}{\bibinfo{person}{Anselm Spoerri}.}
  \bibinfo{year}{1993}\natexlab{}.
\newblock \showarticletitle{InfoCrystal: A Visual Tool for Information
  Retrieval \& Management}. In \bibinfo{booktitle}{\emph{Proceedings of the
  Second International Conference on Information and Knowledge Management}}
  (Washington, D.C., USA) \emph{(\bibinfo{series}{CIKM '93})}.
  \bibinfo{publisher}{Association for Computing Machinery},
  \bibinfo{address}{New York, NY, USA}, \bibinfo{pages}{11–20}.
\newblock
\showISBNx{0897916263}
\urldef\tempurl%
\url{https://doi.org/10.1145/170088.170095}
\showDOI{\tempurl}


\bibitem[Stokes and Hearst(2022)]%
        {stokes2022text}
\bibfield{author}{\bibinfo{person}{Chase Stokes} {and} \bibinfo{person}{Marti
  Hearst}.} \bibinfo{year}{2022}\natexlab{}.
\newblock \bibinfo{title}{Why More Text is (Often) Better: Themes from Reader
  Preferences for Integration of Charts and Text}.
\newblock
\newblock
\showeprint[arxiv]{2209.10789}~[cs.HC]


\bibitem[Stokes et~al\mbox{.}(2023)]%
        {stokes2023striking}
\bibfield{author}{\bibinfo{person}{Chase Stokes}, \bibinfo{person}{Vidya
  Setlur}, \bibinfo{person}{Bridget Cogley}, \bibinfo{person}{Arvind
  Satyanarayan}, {and} \bibinfo{person}{Marti~A. Hearst}.}
  \bibinfo{year}{2023}\natexlab{}.
\newblock \showarticletitle{Striking a Balance: Reader Takeaways and
  Preferences when Integrating Text and Charts}.
\newblock \bibinfo{journal}{\emph{IEEE Transactions on Visualization and
  Computer Graphics}} \bibinfo{volume}{29}, \bibinfo{number}{1}
  (\bibinfo{year}{2023}), \bibinfo{pages}{1233--1243}.
\newblock
\urldef\tempurl%
\url{https://doi.org/10.1109/TVCG.2022.3209383}
\showDOI{\tempurl}


\bibitem[Sumner et~al\mbox{.}(2005)]%
        {linkgoals&resources2005}
\bibfield{author}{\bibinfo{person}{Tamara Sumner}, \bibinfo{person}{Faisal
  Ahmad}, \bibinfo{person}{Sonal Bhushan}, \bibinfo{person}{Qianyi Gu},
  \bibinfo{person}{Francis Molina}, \bibinfo{person}{Stedman Willard},
  \bibinfo{person}{Michael Wright}, \bibinfo{person}{Lynne Davis}, {and}
  \bibinfo{person}{Greg Jan{\'e}e}.} \bibinfo{year}{2005}\natexlab{}.
\newblock \showarticletitle{Linking learning goals and educational resources
  through interactive concept map visualizations}.
\newblock \bibinfo{journal}{\emph{International Journal on Digital Libraries}}
  \bibinfo{volume}{5} (\bibinfo{year}{2005}), \bibinfo{pages}{18--24}.
\newblock


\bibitem[Sutherland(1964)]%
        {sketchpadsutherland1964}
\bibfield{author}{\bibinfo{person}{Ivan~E. Sutherland}.}
  \bibinfo{year}{1964}\natexlab{}.
\newblock \showarticletitle{Sketch Pad a Man-Machine Graphical Communication
  System}. In \bibinfo{booktitle}{\emph{Proceedings of the SHARE Design
  Automation Workshop}} \emph{(\bibinfo{series}{DAC '64})}.
  \bibinfo{publisher}{Association for Computing Machinery},
  \bibinfo{address}{New York, NY, USA}, \bibinfo{pages}{6.329–6.346}.
\newblock
\showISBNx{9781450379328}
\urldef\tempurl%
\url{https://doi.org/10.1145/800265.810742}
\showDOI{\tempurl}


\bibitem[Tang et~al\mbox{.}(2021)]%
        {conceptguide2021}
\bibfield{author}{\bibinfo{person}{Chien-Lin Tang}, \bibinfo{person}{Jingxian
  Liao}, \bibinfo{person}{Hao-Chuan Wang}, \bibinfo{person}{Ching-Ying Sung},
  {and} \bibinfo{person}{Wen-Chieh Lin}.} \bibinfo{year}{2021}\natexlab{}.
\newblock \showarticletitle{ConceptGuide: Supporting Online Video Learning with
  Concept Map-Based Recommendation of Learning Path}. In
  \bibinfo{booktitle}{\emph{Proceedings of the Web Conference 2021}}
  (Ljubljana, Slovenia) \emph{(\bibinfo{series}{WWW '21})}.
  \bibinfo{publisher}{Association for Computing Machinery},
  \bibinfo{address}{New York, NY, USA}, \bibinfo{pages}{2757–2768}.
\newblock
\showISBNx{9781450383127}
\urldef\tempurl%
\url{https://doi.org/10.1145/3442381.3449808}
\showDOI{\tempurl}


\bibitem[Tversky et~al\mbox{.}(2002)]%
        {TVERSKY2002247}
\bibfield{author}{\bibinfo{person}{Barbara Tversky},
  \bibinfo{person}{Julie~Bauer Morrison}, {and} \bibinfo{person}{Mireille
  Betrancourt}.} \bibinfo{year}{2002}\natexlab{}.
\newblock \showarticletitle{Animation: can it facilitate?}
\newblock \bibinfo{journal}{\emph{International journal of human-computer
  studies}} \bibinfo{volume}{57}, \bibinfo{number}{4} (\bibinfo{year}{2002}),
  \bibinfo{pages}{247--262}.
\newblock
\showISSN{1071-5819}
\urldef\tempurl%
\url{https://doi.org/10.1006/ijhc.2002.1017}
\showDOI{\tempurl}


\bibitem[Vaithilingam et~al\mbox{.}(2022)]%
        {evallmcode2022}
\bibfield{author}{\bibinfo{person}{Priyan Vaithilingam},
  \bibinfo{person}{Tianyi Zhang}, {and} \bibinfo{person}{Elena~L. Glassman}.}
  \bibinfo{year}{2022}\natexlab{}.
\newblock \showarticletitle{Expectation vs. Experience: Evaluating the
  Usability of Code Generation Tools Powered by Large Language Models}. In
  \bibinfo{booktitle}{\emph{Extended Abstracts of the 2022 CHI Conference on
  Human Factors in Computing Systems}} (New Orleans, LA, USA)
  \emph{(\bibinfo{series}{CHI EA '22})}. \bibinfo{publisher}{Association for
  Computing Machinery}, \bibinfo{address}{New York, NY, USA}, Article
  \bibinfo{articleno}{332}, \bibinfo{numpages}{7}~pages.
\newblock
\showISBNx{9781450391566}
\urldef\tempurl%
\url{https://doi.org/10.1145/3491101.3519665}
\showDOI{\tempurl}


\bibitem[Victor(2011)]%
        {victor2011abstraction}
\bibfield{author}{\bibinfo{person}{Bret Victor}.}
  \bibinfo{year}{2011}\natexlab{}.
\newblock \bibinfo{title}{Up and Down the Ladder of Abstraction: A Systematic
  Approach to Interactive Visualization}.
\newblock
\newblock
\urldef\tempurl%
\url{http://worrydream.com/LadderOfAbstraction/}
\showURL{%
\tempurl}


\bibitem[Wheeldon(2011)]%
        {wheeldon2011picture}
\bibfield{author}{\bibinfo{person}{Johannes Wheeldon}.}
  \bibinfo{year}{2011}\natexlab{}.
\newblock \showarticletitle{Is a Picture Worth a Thousand Words? Using Mind
  Maps to Facilitate Participant Recall in Qualitative Research.}
\newblock \bibinfo{journal}{\emph{Qualitative Report}} \bibinfo{volume}{16},
  \bibinfo{number}{2} (\bibinfo{year}{2011}), \bibinfo{pages}{509--522}.
\newblock


\bibitem[Winograd et~al\mbox{.}(1972)]%
        {winograd1972shrdlu}
\bibfield{author}{\bibinfo{person}{Terry Winograd} {et~al\mbox{.}}}
  \bibinfo{year}{1972}\natexlab{}.
\newblock \bibinfo{title}{Shrdlu: A system for dialog}.
\newblock
\newblock


\bibitem[Wu et~al\mbox{.}(2012)]%
        {wu2012innovative}
\bibfield{author}{\bibinfo{person}{Po-Han Wu}, \bibinfo{person}{Gwo-Jen Hwang},
  \bibinfo{person}{Marcelo Milrad}, \bibinfo{person}{Hui-Ru Ke}, {and}
  \bibinfo{person}{Yueh-Min Huang}.} \bibinfo{year}{2012}\natexlab{}.
\newblock \showarticletitle{An innovative concept map approach for improving
  students' learning performance with an instant feedback mechanism}.
\newblock \bibinfo{journal}{\emph{British Journal of Educational Technology}}
  \bibinfo{volume}{43}, \bibinfo{number}{2} (\bibinfo{year}{2012}),
  \bibinfo{pages}{217--232}.
\newblock


\bibitem[Wu et~al\mbox{.}(2022a)]%
        {wu2022promptchainer}
\bibfield{author}{\bibinfo{person}{Tongshuang Wu}, \bibinfo{person}{Ellen
  Jiang}, \bibinfo{person}{Aaron Donsbach}, \bibinfo{person}{Jeff Gray},
  \bibinfo{person}{Alejandra Molina}, \bibinfo{person}{Michael Terry}, {and}
  \bibinfo{person}{Carrie~J Cai}.} \bibinfo{year}{2022}\natexlab{a}.
\newblock \showarticletitle{PromptChainer: Chaining Large Language Model
  Prompts through Visual Programming}. In \bibinfo{booktitle}{\emph{Extended
  Abstracts of the 2022 CHI Conference on Human Factors in Computing Systems}}
  (New Orleans, LA, USA) \emph{(\bibinfo{series}{CHI EA '22})}.
  \bibinfo{publisher}{Association for Computing Machinery},
  \bibinfo{address}{New York, NY, USA}, Article \bibinfo{articleno}{359},
  \bibinfo{numpages}{10}~pages.
\newblock
\showISBNx{9781450391566}
\urldef\tempurl%
\url{https://doi.org/10.1145/3491101.3519729}
\showDOI{\tempurl}


\bibitem[Wu et~al\mbox{.}(2022b)]%
        {wu2022ai}
\bibfield{author}{\bibinfo{person}{Tongshuang Wu}, \bibinfo{person}{Michael
  Terry}, {and} \bibinfo{person}{Carrie~Jun Cai}.}
  \bibinfo{year}{2022}\natexlab{b}.
\newblock \showarticletitle{AI Chains: Transparent and Controllable Human-AI
  Interaction by Chaining Large Language Model Prompts}. In
  \bibinfo{booktitle}{\emph{Proceedings of the 2022 CHI Conference on Human
  Factors in Computing Systems}} (New Orleans, LA, USA)
  \emph{(\bibinfo{series}{CHI '22})}. \bibinfo{publisher}{Association for
  Computing Machinery}, \bibinfo{address}{New York, NY, USA}, Article
  \bibinfo{articleno}{385}, \bibinfo{numpages}{22}~pages.
\newblock
\showISBNx{9781450391573}
\urldef\tempurl%
\url{https://doi.org/10.1145/3491102.3517582}
\showDOI{\tempurl}


\bibitem[Xia(2020)]%
        {crosspower2020}
\bibfield{author}{\bibinfo{person}{Haijun Xia}.}
  \bibinfo{year}{2020}\natexlab{}.
\newblock \showarticletitle{Crosspower: Bridging Graphics and Linguistics}. In
  \bibinfo{booktitle}{\emph{Proceedings of the 33rd Annual ACM Symposium on
  User Interface Software and Technology}} (Virtual Event, USA)
  \emph{(\bibinfo{series}{UIST '20})}. \bibinfo{publisher}{Association for
  Computing Machinery}, \bibinfo{address}{New York, NY, USA},
  \bibinfo{pages}{722–734}.
\newblock
\showISBNx{9781450375146}
\urldef\tempurl%
\url{https://doi.org/10.1145/3379337.3415845}
\showDOI{\tempurl}


\bibitem[Xia et~al\mbox{.}(2017)]%
        {writlarge2017}
\bibfield{author}{\bibinfo{person}{Haijun Xia}, \bibinfo{person}{Ken Hinckley},
  \bibinfo{person}{Michel Pahud}, \bibinfo{person}{Xiao Tu}, {and}
  \bibinfo{person}{Bill Buxton}.} \bibinfo{year}{2017}\natexlab{}.
\newblock \showarticletitle{WritLarge: Ink Unleashed by Unified Scope, Action,
  \& Zoom}. In \bibinfo{booktitle}{\emph{Proceedings of the 2017 CHI Conference
  on Human Factors in Computing Systems}} (Denver, Colorado, USA)
  \emph{(\bibinfo{series}{CHI '17})}. \bibinfo{publisher}{Association for
  Computing Machinery}, \bibinfo{address}{New York, NY, USA},
  \bibinfo{pages}{3227–3240}.
\newblock
\showISBNx{9781450346559}
\urldef\tempurl%
\url{https://doi.org/10.1145/3025453.3025664}
\showDOI{\tempurl}


\bibitem[Xia et~al\mbox{.}(2020)]%
        {crosscast2020}
\bibfield{author}{\bibinfo{person}{Haijun Xia}, \bibinfo{person}{Jennifer
  Jacobs}, {and} \bibinfo{person}{Maneesh Agrawala}.}
  \bibinfo{year}{2020}\natexlab{}.
\newblock \showarticletitle{Crosscast: Adding Visuals to Audio Travel
  Podcasts}. In \bibinfo{booktitle}{\emph{Proceedings of the 33rd Annual ACM
  Symposium on User Interface Software and Technology}} (Virtual Event, USA)
  \emph{(\bibinfo{series}{UIST '20})}. \bibinfo{publisher}{Association for
  Computing Machinery}, \bibinfo{address}{New York, NY, USA},
  \bibinfo{pages}{735–746}.
\newblock
\showISBNx{9781450375146}
\urldef\tempurl%
\url{https://doi.org/10.1145/3379337.3415882}
\showDOI{\tempurl}


\bibitem[Yuan et~al\mbox{.}(2022)]%
        {yuan2022wordcraft}
\bibfield{author}{\bibinfo{person}{Ann Yuan}, \bibinfo{person}{Andy Coenen},
  \bibinfo{person}{Emily Reif}, {and} \bibinfo{person}{Daphne Ippolito}.}
  \bibinfo{year}{2022}\natexlab{}.
\newblock \showarticletitle{Wordcraft: Story Writing With Large Language
  Models}. In \bibinfo{booktitle}{\emph{27th International Conference on
  Intelligent User Interfaces}} (Helsinki, Finland) \emph{(\bibinfo{series}{IUI
  '22})}. \bibinfo{publisher}{Association for Computing Machinery},
  \bibinfo{address}{New York, NY, USA}, \bibinfo{pages}{841–852}.
\newblock
\showISBNx{9781450391443}
\urldef\tempurl%
\url{https://doi.org/10.1145/3490099.3511105}
\showDOI{\tempurl}


\bibitem[Zhang et~al\mbox{.}(2023)]%
        {zhang2023language}
\bibfield{author}{\bibinfo{person}{Muru Zhang}, \bibinfo{person}{Ofir Press},
  \bibinfo{person}{William Merrill}, \bibinfo{person}{Alisa Liu}, {and}
  \bibinfo{person}{Noah~A. Smith}.} \bibinfo{year}{2023}\natexlab{}.
\newblock \bibinfo{title}{How Language Model Hallucinations Can Snowball}.
\newblock
\newblock
\showeprint[arxiv]{2305.13534}~[cs.CL]


\bibitem[Zhou et~al\mbox{.}(2022)]%
        {promptingvizlang2022}
\bibfield{author}{\bibinfo{person}{Kaiyang Zhou}, \bibinfo{person}{Jingkang
  Yang}, \bibinfo{person}{Chen~Change Loy}, {and} \bibinfo{person}{Ziwei Liu}.}
  \bibinfo{year}{2022}\natexlab{}.
\newblock \showarticletitle{Learning to prompt for vision-language models}.
\newblock \bibinfo{journal}{\emph{International Journal of Computer Vision}}
  \bibinfo{volume}{130}, \bibinfo{number}{9} (\bibinfo{year}{2022}),
  \bibinfo{pages}{2337--2348}.
\newblock


\end{thebibliography}

%%
%% Appendix, this is the place to put it.
\appendix

\appendix

\section{Prompts}
\label{app:p}

\rv{We provide all the prompts that we use for the prototype environment of \mbox{\system} with OpenAI GPT-4 API.} \textbf{System}, \textbf{User}, and \textbf{Assistant} (i.e. GPT-4) are pre-defined roles for querying the API\footnote{https://platform.openai.com/docs/api-reference/chat}.

\subsection{Initial Query}
\label{app:p-initial}

\noindent\textbf{System} Please provide a well-structured response to the user's question in multiple paragraphs. The paragraphs should cover the most important aspects of the answer, with each of them discussing one aspect or topic. Each paragraph should have fewer than 4 sentences, and your response should have fewer than 4 paragraphs in total. The user’s goal is to construct a concept map to visually explain your response. To achieve this, annotate the key entities and relationships inline for each sentence in the paragraphs.

Entities are usually noun phrases and should be annotated with [entity (\$N1)], for example, [Artificial Intelligence (\$N1)]. Do not annotate conjunctive adverbs, such as ``since then'' or ``therefore'', as entities in the map.

A relationship is usually a word or a phrase that consists of verbs, adjectives, adverbs, or prepositions, e.g., ``contribute to'', ``by'', ``is'', and ``such as''. Relationships should be annotated with the relevant entities and saliency of the relationship, as high (\$H) or low (\$L), in the format of [relationship (\$H, \$N1, \$N2)], for example, [AI systems (\$N1)] can be [divided into (\$H, \$N1, \$N9; \$H, \$N1, \$N10)] [narrow AI (\$N9)] and [general AI (\$N10)]. Relationships of high saliency are those included in summaries. Relationships of low saliency are often omitted in summaries. It's important to choose relationships that accurately reflect the nature of the connection between the entities in text, and to use a consistent annotation format throughout the paragraphs.

You should try to annotate at least one relationship for each entity. Relationships should only connect entities that appear in the response. You can arrange the sentences in a way that facilitates the annotation of entities and relationships, but the arrangement should not alter their meaning, and they should still flow naturally in language.

Example paragraph A: [Artificial Intelligence (AI) (\$N1)] [is a (\$H, \$N1, \$N2)] [field of computer science (\$N2)] that [creates (\$H, \$N1, \$N3)] [intelligent machines (\$N3)]. [These machines (\$N3)] [possess (\$H, \$N3, \$N4)] [capabilities (\$N4)] [such as (\$L, \$N4, \$N5; \$L, \$N4, \$N6; \$L, \$N4, \$N7; \$L, \$N4, \$N8)] [learning (\$N5)], [reasoning (\$N6)], [perception (\$N7)], and [problem-solving (\$N8)]. [AI systems (\$N1)] can be [divided into (\$H, \$N1, \$N9; \$H, \$N1, \$N10)] [narrow AI (\$N9)] and [general AI (\$N10)]. [Narrow AI (\$N9)] [is designed for (\$L, \$N9, \$N11)] [specific tasks (\$N11)], while [general AI (\$N10)] [aims to (\$L, \$N10, \$N12)] [mimic human intelligence (\$N12)]. [It (\$N1)] [has grown across (\$H, \$N1, \$N13)] [multiple industries (\$N13)], [leading to (\$L, \$N1, \$N14; \$L, \$N1, \$N15; \$L, \$N1, \$N16)] [improved efficiency (\$N14)], [enhanced decision-making (\$N15)], and [better user experiences (\$N16)].

Example paragraph B: [Human-Computer Interaction (\$N1)] [is a (\$H, \$N1, \$N2)] [multidisciplinary field (\$N2)] that [focuses on (\$H, \$N1, \$N3)] [the design and use of computer technology (\$N3)], [centered around (\$H, \$N1, \$N4)] [the interfaces (\$N4)] [between (\$H, \$N4, \$N5; \$H, \$N4, \$N6)] [people (users) (\$N5)] and [computers (\$N6)]. [Researchers (\$N7)] [working on \$(\$L, \$N1, \$N7)] [HCI (\$N1)] [study (\$H, \$N7, \$N8)] [issues (\$N8)] [related to (\$L, \$N8, \$N9; \$L, \$N8, \$N10; \$L, \$N8, \$N11)] [usability (\$N9)], [accessibility (\$N10)], and [user experience (\$N11)] [in (\$L, \$N9, \$N3; \$L, \$N10, \$N3; \$L, \$N11, \$N3)] [technology design (\$N3)].

Example paragraph C: [Birds (\$N1)] [can (\$H, \$N1, \$N2)] [fly (\$N2)] [due to (\$H, \$N2, \$N3)] [a combination of physiological adaptations (\$N3)]. [One key (\$H, \$N3, \$N4)] [adaptation (\$N4)] [is (\$H, \$N4, \$N5)] the [presence of lightweight bones (\$N5)] that [reduce (\$H, \$N5, \$N6)] [their body weight (\$N6)], [making (\$L, \$N5, \$N7)] it [easier for them to fly (\$N7)]. [Another (\$H, \$N3, \$N8)] [adaptation (\$N8)] [is (\$H, \$N8, \$N9)] the [structure of their wings (\$N9)] which [are designed for (\$H, \$N9, \$N2)] [flight (\$N2)].

Your response should have multiple paragraphs.

\noindent\textbf{User} \textit{[The query provided by the user.]}

\subsection{Self-Correction}
\label{app:p-corr}

\noindent\textit{[The initial prompts and responses from the \textbf{System}, \textbf{User}, and \textbf{Assistant}.]}

\noindent\textbf{System} In the following sentence of your original response, there are some issues that need to be fixed.

The entities \textit{[list the annotated entities, separate by commas]} were mentioned but not connected by any relationships. \textit{[Add this paragraph only if orphan nodes were detected.]}

One or more relationships annotated by relationship annotations \textit{[list the annotated relationships, separate by commas]} were trying to connect entities with ids that are not mentioned in the response. \textit{[Add this paragraph only if dead-end relationships were detected.]}

In your corrected response, please make sure that all entities and relationships are extracted correctly. Relationships should only connect existing entities, and entities should be connected by at least one relationship. Please try to fix these issues in your response by annotating the same sentence again. You may arrange the sentences in a way that facilitates the annotation of entities and relationships, but the arrangement should not alter their meaning and they should still flow naturally in language.

When annotating a new entity that was not mentioned in the previous response, please make sure that they are annotated with a new entity id (for example, if the previous annotation has reached id ``\$N102'', then the new annotation id should start at ``\$N103''). However, if the same entity has appeared in the original response, please match their id.

Please only include the re-annotated sentence in your response.

\subsection{Summary}
\label{app:p-sum}

\noindent\textbf{System} You are a professional writer specializing in text summarization. Make a short, one-sentence summary of the chunk of the text provided by the user. The summary should reflect the main idea and the most important relationships of the text. Notice that the user has annotated the text with entities and relationships. Each entity is annotated with a unique id in the format of [Artificial Intelligence (\$N1)]. Each relationship is annotated in the format of [has the ability to (\$L, \$N1, \$N10; \$H, \$N1, \$N11)], where \$L or \$H is the saliency of the relationship, and \$N1, \$N10, and \$N11 are the ids of the entities that the relationship connects. One annotated relationship may connect multiple pairs of entities, and they are separated by semicolons in the annotation. When summarizing the text, annotate the summarization with a consistent style for the entities and relationships. Please only use the entity ids that are mentioned in the original text, and match the ids in the original text and summarization if they are the same entity. Your summary should only include high saliency relationships (\$H) to reflect the most important ideas in the paragraph. You can arrange the sentences in the summarization in a way that facilitates the annotation of entities and relationships, but the arrangement should not alter their meaning and they should still flow naturally in language. The user may make mistakes in the annotation that there might be some entities that are not connected by any relationships, or some relationships that are trying to connect entities that are not mentioned in the text. Please avoid these mistakes when annotating the summary. Your summary should have only one short sentence.

Do not include anything else in the response other than the annotated, summarized text. For example, for paragraph:
[Human-Computer Interaction (\$N1)] [is a (\$H, \$N1, \$N2)] [multidisciplinary field (\$N2)] that [focuses on (\$H, \$N1, \$N3)] [the design and use of computer technology (\$N3)], [centered around (\$H, \$N1, \$N4)] [the interfaces (\$N4)] [between (\$H, \$N4, \$N5; \$H, \$N4, \$N6)] [people (users) (\$N5)] and [computers (\$N6)]. [Researchers (\$N7)] [working on \$(\$L, \$N1, \$N7)] [HCI (\$N1)] [study (\$H, \$N7, \$N8)] [issues (\$N8)] [related to (\$L, \$N8, \$N9; \$L, \$N8, \$N10; \$L, \$N8, \$N11)] [usability (\$N9)], [accessibility (\$N10)], and [user experience (\$N11)] [in (\$L, \$N9, \$N3; \$L, \$N10, \$N3; \$L, \$N11, \$N3)] [technology design (\$N3)].

You may summarize it as:
[HCI (\$N1)] [is a (\$H, \$N1, \$N2)] [multidisciplinary field (\$N2)] that [centered around (\$H, \$N1, \$N4)] [the interfaces (\$N4)] [between (\$H, \$N4, \$N5; \$H, \$N4, \$N6)] [users (\$N5)] and [computers (\$N6)].

\noindent\textbf{User} \textit{[The paragraph to be summarized, from the original response.]}

\subsection{Outline}
\label{app:p-outline}

\noindent\textbf{System} You are a professional presentation slide builder. Structure the following text provided by the user into a presentation slide, in markdown format. If you need to use a list, use a numbered list. Do not include anything else in the response other than the markdown text.

\noindent\textbf{User} \textit{[The paragraph to create the outline, from the original response.]}

\subsection{Node Explanation}

\noindent\textit{[The initial prompts and responses from the \textbf{System}, \textbf{User}, and \textbf{Assistant}.]}

\noindent\textbf{User} In the sentence \textit{[the sentence containing the node from the original response]}, you mentioned the entity \textit{[node label]}. Can you explain this entity in 1 to 2 sentences? Please refer to the original response as the context of your explanation. Your explanation should be concise, one paragraph, and follow the same annotation format as the original response. You should try to annotate at least one relationship for each entity. Relationships should only connect entities that appear in the response. When annotating a new entity that was not mentioned in the previous response, please make sure that they are annotated with a new entity id (for example, if the previous annotation has reached id ``\$N102'', then the new annotation id should start at ``\$N103''). However, if the same entity has appeared in the original response, please match their id.

For example, for ``[general AI (\$N10)]'' in the sentence ``[AI systems (\$N1)] can be [divided into (\$H, \$N1, \$N9; \$H, \$N1, \$N10)] [narrow AI (\$N9)] and [general AI (\$N10)].'':
[General AI (\$N10)] refers to a [type of (\$L, \$N1, \$N10)] [artificial intelligence (\$N1)] that [has the ability to (\$L, \$N10, \$N14; \$L, \$N10, \$N5; \$L, \$N10, \$N15)] [understand (\$N14)], [learn (\$N5)], and [apply knowledge across a wide range of tasks (\$N15)].

\balance

\subsection{Node Examples}

\noindent\textit{[The initial prompts and responses from the \textbf{System}, \textbf{User}, and \textbf{Assistant}.]}

\noindent\textbf{User} In the sentence \textit{[the sentence containing the node from the original response]}, you mentioned the entity \textit{[node label]}. Can you give a few examples of it? Your response should follow the same annotation format as the original response, as shown in the following example. When annotating a new entity that was not mentioned in the previous response, please make sure that they are annotated with a new entity id (for example, if the previous annotation has reached id ``\$N102'', then the new annotation id should start at ``\$N103''). However, if the same entity has appeared in the original response, please match their id. You don't need to further explain the examples you give.

For example, for ``[Fruits (\$N1)]'' in the sentence ``[Fruits (\$N1)] can [help with (\$H, \$N1, \$N2)] [health (\$N2)].'', your response could be:
``[Fruits (\$N1)], for example, [includes (\$H, \$N1, \$N3; \$H, \$N1, \$N4; \$H, \$N1, \$N5)], [apples (\$N3)], [oranges (\$N4)], and [watermelons (\$N5)].''

\subsection{Tell Me More}

\noindent\textit{[The initial prompts and responses from the \textbf{System}, \textbf{User}, and \textbf{Assistant}.]}

\noindent\textbf{User} For the paragraph \textit{[the paragraph to be extended]}, can you continue writing one or two more sentences at the end of the paragraph? When continue writing this paragraph, please refer to the original response as the context of your writing. Your response should be about the same topic and aspect of the original paragraph and could add more details. Your response should follow the same annotation format as the original response. When annotating a new entity that was not mentioned in the previous response, please make sure that they are annotated with a new entity id (for example, if the previous annotation has reached id ``\$N102'', then the new annotation id should start at ``\$N103''). However, if the same entity has appeared in the original response, please match their id. Your response should only have the new content.

\subsection{Add a Paragraph}

\noindent\textit{[The initial prompts and responses from the \textbf{System}, \textbf{User}, and \textbf{Assistant}.]}

\noindent\textbf{User} Can you continue writing one paragraph after the end of your original response? When writing the new paragraph, please refer to the original response as the context of your writing. Your response should still try to answer the user's original question and could add more details or provide a new aspect. Your response should follow the same annotation format as the original response. When annotating a new entity that was not mentioned in the previous response, please make sure that they are annotated with a new entity id (for example, if the previous annotation has reached id ``\$N102'', then the new annotation id should start at ``\$N103''). However, if the same entity has appeared in the original response, please match their id. Your response should only have the new content.

\onecolumn
\section{Technical Evaluation Explanation and Examples}
\label{app:ex}

% \usepackage{multirow}
\begin{table*}[htbp]
\centering
\caption{Annotation Error Explanation and Examples}
\label{tab:app-table}
\begin{tabular}{ccp{11cm}}
\toprule
\textbf{Category} & \textbf{Error} & \multicolumn{1}{c}{\textbf{Content}} \\ \midrule
\multirow{8}{*}[-18ex]{\textbf{\begin{tabular}[c]{@{}c@{}}Node\\ Annotation\\ Error Types\end{tabular}}} & \textbf{Missing Entity Phrase} & An entity that is not annotated when should. \\
 & Example & ... [is driven by (\$H, \$N8, \$N9){]} the {[}need to reduce (\$H, \$N7, \$N10)] [greenhouse gas emissions (\$N10){]} from \textbf{traditional internal combustion engines}. \\ \\ \cline{2-3} \\
 & \textbf{Incomplete Entity} & Entity annotations lacking necessary label words or splitting a single entity phrase. \\
 & Example & {[}Impressionist painters (\$N1)] [belong to (\$H, \$N1, \$N2){]} \textbf{{[}the Impressionism (\$N2)] [art movement (\$N3){]}}. \\ \\ \cline{2-3} \\
 & \textbf{Incorrect Entity} & An entity annotation that includes words or phrases that do not belong to the entity. \\
 & Example & \textbf{{[}However (\$N13)]}, [the Industrial Revolution (\$N13){]} {[}led to (\$H, \$N13, \$N14)] [mass production (\$N14){]} of {[}clothing (\$N3){]}. \\ \\ \cline{2-3} \\ 
 & \textbf{Incorrect Co-reference} & Labeling co-referenced entities with different identifiers. \\
 & Example & \textbf{{[}Apple (\$N1)]} [is (\$H, \$N1, \$N2){]} {[}good (\$N2)]. \textbf{[It (\$N3){]}} {[}helps (\$H, \$N3, \$N4)] [health (\$N4){]}. \textit{(Count as 1 incorrect.)} \\ \\ \hline\hline \\
\multirow{10}{*}[-17ex]{\textbf{\begin{tabular}[c]{@{}c@{}}Relationship\\ Annotation\\ Error Types\end{tabular}}} & \textbf{Missing Relationship} & A relationship that is not annotated when should. \\
 & Example & ... \textbf{{[}establishing (\$H, \$N1, \$N5)]} [political systems (\$N5){]}, {[}economies (\$N6)], and [cultural practices (\$N7){]}. \textit{(Missing relationships for economies and cultural practices.)} \\ \\ \cline{2-3} \\ 
 & \textbf{Incomplete Relationship} & A relationship annotation that does not include the whole phrase as the label. \\
 & Example & {[}should be (\$H, \$N12, \$N13)] [based on (\$H, \$N13, \$N14){]} \\ \\ \cline{2-3} \\ 
 & \textbf{Dead-end Relationship} & A relationship annotation that includes entity identifiers that do not exist. \\
 & Example & {[}These philosophers (\$N9)] \textbf{[emphasized (\$H, \$N9, \$N13){]} {[}the importance of (\$L, \$N13, \$N14; \$L, \$N13, \$N15){]}} {[}subjectivity (\$N14)] and [individual freedom (\$N15){]}. \\ \\ \cline{2-3} \\ 
 & \textbf{Reversed Relationship} & A relationship that has been annotated with reversed direction. \\
 & Example & {[}Apple (\$N1)] \textbf{[is (\$H, \$N2, \$N1){]}} {[}good (\$N2){]}. \\ \\ \cline{2-3} \\ 
 & \textbf{Misattributed Relationship} & A relationship that is annotated with the wrong pair of entities. \\
 & Example & {[}Gods (\$N4)] in [Greek mythology (\$N1){]} {[}play (\$H, \textbf{\$N1, \$N4})] [central roles (\$N10){]} in {[}the narratives (\$N11)]. \textit{(Should be \$N4, \$N10.)} \\ \\ \hline\hline \\
\multirow{2}{*}[-5ex]{\textbf{\begin{tabular}[c]{@{}c@{}}Detectable\\ Error Types\end{tabular}}} & \textbf{Orphan Node} & Nodes that are not involved in any relationships, might be caused by Incorrect Entity or Incomplete Relationship. (Percentage divided by Total Extracted Entity Phrase.) \\ \\ \cline{2-3} \\ 
 & \textbf{Dead-end Relationship} & Links that try to connect non-existent nodes, caused by Dead-end Relationship. (Percentage divided by Total Extracted Relationship.) \\ \bottomrule
\end{tabular}
\end{table*}

% \received{18 May 2023}

\end{document}